\documentclass[aps,pre,preprint,onecolumn,nofootinbib,superscriptaddress]{revtex4-1}

\usepackage{graphicx}
\usepackage{natbib}
\usepackage{amsmath}
\usepackage{amssymb}
\usepackage[nointegrals]{wasysym}
\usepackage{color}
\usepackage[T1]{fontenc}
\usepackage[colorlinks]{hyperref}

\makeatletter
\newcommand{\rmnum}[1]{\romannumeral #1}
\newcommand{\Rmnum}[1]{\expandafter\@slowromancap\romannumeral #1@}
\makeatother

\begin{document}

\title{Revisiting the Anomalous Bending Elasticity of Sharply Bent DNA}

\author{Peiwen Cong}
\affiliation{Mechanobiology Institute, National University of Singapore, Singapore 117411, Singapore}
\affiliation{Computation and Systems Biology, Singapore-MIT Alliance, Singapore 117576, Singapore}
\affiliation{Department of Physics, National University of Singapore, Singapore 117551, Singapore}
\author{Liang Dai}
\affiliation{BioSystems and Micromechanics IRG, Singapore-MIT Alliance for Research and Technology Centre, Singapore 138602, Singapore}
\author{Hu Chen}
\affiliation{Department of Physics, Xiamen University, Xiamen, Fujian 361005, China}
\author{Johan R. C. van der Maarel}
\affiliation{Department of Physics, National University of Singapore, Singapore 117551, Singapore}
\author{Patrick S. Doyle}
\affiliation{Department of Chemical Engineering, Massachusetts Institute of Technology, Cambridge, Massachusetts 02139, USA}
\affiliation{BioSystems and Micromechanics, Singapore-MIT Alliance for Research and Technology Centre, Singapore 138602, Singapore}
\author{Jie Yan}
\email[Email address: ]{phyyj@nus.edu.sg}
\affiliation{Department of Physics, National University of Singapore, Singapore 117551, Singapore}
\affiliation{Mechanobiology Institute, National University of Singapore, Singapore 117411, Singapore}
\affiliation{Centre for BioImaging Sciences, National University of Singapore, Singapore 117557, Singapore}
\affiliation{BioSystems and Micromechanics, Singapore-MIT Alliance for Research and Technology Centre, Singapore 138602, Singapore}

\begin{abstract}
Several recent experiments suggest that sharply bent DNA has a surprisingly
high bending flexibility, but the cause of this flexibility is poorly
understood. Although excitation of flexible defects can explain these results,
whether such excitation can occur with the level of DNA bending in these
experiments remains unclear. Intriguingly, the DNA contained preexisting nicks
in most of these experiments but whether nicks might play a role in flexibility
has never been considered in the interpretation of experimental results. Here,
using full-atom molecular dynamics simulations, we show that nicks promote DNA
basepair disruption at the nicked sites, which drastically reduces DNA bending
energy. In addition, lower temperatures suppress the nick-dependent basepair
disruption. In the absence of nicks, basepair disruption can also occur but
requires a higher level of DNA bending. Therefore, basepair disruption inside
\textit{B}-form DNA can be suppressed if the DNA contains preexisting nicks.
Overall, our results suggest that the reported mechanical anomaly of sharply
bent DNA is likely dependent on preexisting nicks, therefore the intrinsic
mechanisms of sharply bent nick-free DNA remain an open question.
\end{abstract}

\maketitle

\section{INTRODUCTION}

Many cellular processes such as DNA packaging and gene transcription require
sharp DNA bending \cite{Richmond:2003hm, Oehler:1994vz}. Thus, knowledge of the
mechanics of sharply bent DNA is critical to understand these cellular
processes. DNA is often modeled as a linear polymer that is described by a
spatial curve in three dimensions. The bending rigidity of nonsharply bent DNA
has been described by the wormlike chain (WLC) polymer model
\cite{Hagerman:1988}. In the WLC polymer model, the bending energy of short DNA
is described by $\beta E \left( \theta; A \right) = \left(A / 2L \right) \left(
\hat{t}^\prime - \hat{t} \right)^2= \left( A / L \right) \left( 1-\cos\theta
\right)$, where $A$ is the bending persistence length of DNA. Here $\beta = 1 /
{k_\text{B}T}$ scales energy into units of $k_\text{B}T$; $L \ll A$ is the DNA
contour length; $\hat{t}$, $\hat{t}^\prime$ are the tangent vectors at two DNA
ends; and $\theta$ is the bending angle of DNA. The bending persistence length
of \textit{B}-form DNA has been experimentally determined to be $A \approx 50$
nm \cite{Shore:1981wt, Doi:1986ug, Smith:1992em, Marko:1995}. This bending
rigidity is also related to the Young's elasticity modulus $Y$ of elastic rods
through the equation $A=\beta Y I$. Here $I = \pi R^4 / 4$ is the DNA area
moments of inertia, while $R$ is its radius.

The mechanical anomaly of sharply bent DNA was reported in several recent
experiments. In particular, the probabilities of spontaneous looping of $\sim
100$ bp DNA into minicircles were several orders of magnitude larger than
predicted by the WLC model \cite{Cloutier:2004fg, Vafabakhsh:2012dt}. The level
of DNA bending in such DNA minicircles is biologically relevant given its
similar level of bending to DNA wrapping around nucleosomes \cite{Luger:1997gf,
Davey:2002ir}. While the mechanical anomaly of sharply bent DNA has drawn
extensive interest, the underlying mechanisms remain unclear and debated. This
work aims to provide insights into this debate using full-atom molecular
dynamics (MD) simulations. To help readers understand the question and the
motivation of this work, we first review previous DNA looping experiments and
the underlying assumptions used to interpret those results.

\subsection*{Jacobson-Stockmayer factor measurements}

The debate surrounding the mechanisms of sharply bent DNA flexibility began
with a Jacobson-Stockmayer factor ($j$-factor) measurement
\cite{Jacobson:1950kv}, which reported an anomalously high probability of DNA
looping at $94-116$ bp \cite{Cloutier:2004fg}. These experiments used a DNA
fragment with short strands of complementary single-stranded DNA (ssDNA)
overhanging on each end. In a solution at a concentration $c=N / {V}$ ($N$ is
the number of molecules and $V$ is the volume), a terminal end can hybridize
with the complementary end on the same DNA fragment (i.e., looping) or with the
end of a different DNA fragment (i.e., dimerization), which is driven by
thermal fluctuation. Theoretically, the ratio of the looping and dimerization
rates predicts the probability density of spontaneous looping in competition
with hybridization to a nearby DNA molecule. This probability density is
determined by $j$-factor measurements with the following equation:
$\rho^{\text{J}} \! \left(\boldsymbol{0} \right)=c^\prime \times {\left(
r_{\text{loop}} / r_{\text{dimer}} \right)} = {\left( c^\prime / c \right)} {
\left( r_{\text{loop}} / r^0_{\text{dimer}} \right)}$. In this equation,
$\boldsymbol{0}$ indicates zero end-to-end distance vector, $c^\prime < c$ is
the concentration of DNA fragments with orientations allowing for
hybridization, and $r^0_{\text{dimer}}=r_{\text{dimer}}/c$ denotes the
dimerization rate per unit concentration of DNA. The superscript ``J''
indicates that $\rho^{\text{J}} \! \left(\boldsymbol{0} \right)$ is determined
by $j$-factor measurements.

According to this equation, the looping probability can be experimentally
determined from the ratio of looping and dimerization rates, which can be
measured by chemically fixing the populations of looped and dimerized DNA
species with a ligation reaction \cite{Shore:1981wt, Shore:1983jm}.
Importantly, equilibration of the double-nicked DNA intermediates (looped
fragments and dimers) prior to ligation is a prerequisite. In other words,
$j$-factor measurements probe the looping probability of a subset of
double-nicked looped DNA intermediates that can be covalently sealed by ligase
(see Discussion).

A $j$-factor with units of concentration is often defined as $j=r_{\text{loop}}
/ r^0_{\text{dimer}}$ \cite{Shore:1981wt, Cloutier:2004fg, Du:2005gg};
therefore $\rho^{\text{J}} \! \left(\boldsymbol{0} \right)={\left( c^\prime / c
\right)} \times j$. To calculate $\rho^{\text{J}} \! \left(\boldsymbol{0}
\right)$ from $j$, prior knowledge of the ${c^\prime / c}$ is needed. It is
known that a nick on a linear DNA does not affect basepairing and stacking at
the nicked site; therefore, hybridized DNA ends in dimerized linear DNA are in
parallel and twist-matching to each other to form the \textit{B}-DNA
conformation \cite{ClausRoll:1998dl, Hyz:2011dv}. Hereafter we refer to this
constraint as ``twist-matching parallel boundary condition'', denoted by
``$\Omega$'' (Fig.~\ref{fig:1Orientation}). This results in ${c^\prime / c} =
{\left( 4 \pi \times 2 \pi \right) }^{-1}$, where $4 \pi$ arises from the
constraint for tangential parallel alignment, while $2 \pi$ comes from
twist-matching for the dimerization reaction and thus results in
$\rho^{\text{J}} \! \left( \boldsymbol{0} \right)= j / {\left(8\pi^2 \right)}$.

To draw information of the elasticity of DNA bending from the measured DNA
looping probability density in these $j$-factor measurements, $\rho^{\text{J}}
\! \left( \boldsymbol{0} \right)$ can be compared with the theoretical looping
probability density $\rho^{\text{WLC}}_{\xi} \left( \boldsymbol{0} \right)$.
This is based on the WLC model according to an appropriate constraint ($\xi$),
on the orientations of the two ends in the looped DNA. In previous studies,
$\xi$ has been assumed to be $\Omega$, which is the same as that imposed on
dimerized DNA. Based on $\rho^{\text{WLC}}_{\Omega} \left( \boldsymbol{0}
\right)=\rho^{\text{J}} \! \left( \boldsymbol{0} \right)$, the DNA persistence
length was determined to be in the range of $45-55$ nm, over a wide contour
length ($> 200$ bp) in normal solution conditions \cite{Shore:1983jm,
Taylor:1990ve}. The agreement of the persistence length $A$ determined in
$j$-factor measurements and that determined in single-DNA stretching
experiments validates the $\Omega$-boundary condition for both looped and
dimerized DNA with sizes larger than $200$ bp.

However, for shorter DNA fragments at $\sim 100$ bp, $\rho^{\text{J}} \! \left(
\boldsymbol{0} \right)$ is several orders of magnitude larger than
$\rho^{\text{WLC}}_{\Omega} \left( \boldsymbol{0} \right)$ predicted with $A
\approx 50$ nm \cite{Cloutier:2004fg, Cloutier:2005et}. There are two possible
causes of such discrepancy: (\textit{i}) an intrinsic elastic response of
double-stranded DNA (dsDNA) under sharp bending condition might occur by
bending-induced flexible defects excited inside the DNA as proposed by several
groups \cite{Cloutier:2004fg, Yan:2004vt, Yan:2005vk, Wiggins:2005uu,
Destainville:2009fw}; and (\textit{ii}) the $\Omega$-boundary condition
assumption is no longer valid for the hybridized looped DNA when DNA is sharply
bent. Violation of the $\Omega$-boundary condition assumption could occur if
the nicked sites on two hybridized ends on a sharply bent DNA loop could not
maintain the \textit{B}-form conformation. This possibility has not been
considered to interpret the apparent disagreement between $\rho^{\text{J}} \!
\left( \boldsymbol{0} \right)$ and $\rho^{\text{WLC}}_{\Omega} \left(
\boldsymbol{0} \right)$ in previous $j$-factor studies.

\subsection*{Single-molecule F\"{o}rster resonance energy transfer experiments}

The mechanical anomaly of sharply bent DNA was also reported in two recent
studies that employed single-molecule F\"{o}rster resonance energy transfer
(smFRET) \cite{Vafabakhsh:2012dt, Le:2014iw}. In these studies, complimentary
ssDNA overhangs at each end of a short DNA fragment were used to stabilize the
looped conformation to achieve a sufficiently long lifetime needed for smFRET
measurements. Therefore, this looped DNA contained two nicks, which is similar
to the looped DNA in the $j$-factor measurement before the ligation reactions.

In the first study, the looping probability was determined as a measure of the
lifetimes of the looped and unlooped DNA \cite{Vafabakhsh:2012dt}. Similar to
the $j$-factor measurement, an anomalously high looping probability was
observed for DNA at $\sim 100$ bp compared to that predicted with the WLC model
using the $\Omega$-boundary condition. In the second study \cite{Le:2014iw},
the relationship of loop lifetime and the bending stress analyzed in
$\Omega$-boundary condition also revealed anomalous DNA bending elasticity for
DNA fragments $< 100$ bp \cite{Le:2014iw}. However, considering the presence of
nicks in the hybridized DNA loops, these experiments could also be explained by
a violation of the $\Omega$-boundary condition at the nicked sites.

In summary of these DNA looping experiments, the DNA contained preexisting
nicks. It is generally assumed that nicks do not affect the local mechanical
properties of sharply bent DNA, thus the observed mechanical anomaly can be
explained by a breakdown of the WLC polymer model. Indeed, it has been
theoretically predicted that excitation of flexible mechanical defects under
bending constraints by way of DNA melting or kinking can explain these results
\cite{Yan:2004vt, Yan:2005vk, Wiggins:2005uu}. On the other hand, as we
mentioned, the mechanical anomaly of sharply bent DNA could also be explained
by violation of the $\Omega$-boundary condition at the nicked sites.

The potential role of nicks in the DNA looping assays was only mentioned as a
possible cause of the apparent DNA mechanical anomaly \cite{Chen:2008ed,
Vologodskii:2013ca}; however, whether a nick can promote excitation of a
mechanical defect at the nicked site has never been quantitatively
investigated. Under sharp bending constraints, it is possible that the nicked
site might unstack, causing the formation of a flexible defect that reduces the
overall bending energy of the looped DNA. As such, defect excitation would not
occur in the nick-free region of DNA due to the relaxed bending in the
nick-free region because of flexible defect excitation at the nicks.

In this work, we carried out full-atom MD simulations to investigate the
mechanical responses of short dsDNA fragment ($20$ bp) under compressive load
in the absence and presence of a nick in the DNA (see Materials and Methods for
details on DNA constructs, spring constraints, and MD simulations).

We show that sharp DNA bending that is induced using sufficiently stiff springs
with zero equilibrium length leads to local DNA basepair disruptions.
Subsequently, DNA kinks with large bending angles develop around the disrupted
DNA basepairs, which relaxes the bending of the rest of DNA. We also
demonstrate that a nick is a structurally weaker point than basepairs in a
nick-free DNA region. Thus, under sharp bending conditions nicks often lead to
unstacked (basepairs intact) or peeled (basepair-disrupted) DNA, resulting in
DNA kink formation localized to the nicked site. Furthermore, this
nick-dependent defect excitation is sensitive to temperature changes within a
physiological range.

In summary, nicks promote flexible defect excitation under sharp bending
constraints, resulting in the formation of a DNA kink localized at the nicked
site, which in turn prevents defect excitation in the nick-free DNA region.
Based on these results, we suggest that the previously reported mechanical
anomaly of sharply bent DNA can alternatively be explained as being
attributable to nick-dependent flexible defect excitation.

In the Materials and Methods, we provide concise information about:
(\textit{i}) DNA constructs; (\textit{ii}) spring constraints for generating
sharp DNA bending and for umbrella sampling analysis; and (\textit{iii}) force
field, water model, software, and other simulation aspects. In the Results, we
show what is obtained on sharply bent nick-free DNA. We then present the free
energy landscape and the force needed to maintain certain end-to-end distance
obtained using umbrella sampling, for nick-free DNA before and after
disruptions of basepairs. We also present the results of the nick-dependent
defect excitation in sharply bent nick-containing DNA. In the Discussion, we
provide the implications of these findings in relation to the reported
anomalous DNA bending elasticity of sharply bent DNA molecules.

\section{MATERIALS AND METHODS}
\subsection*{DNA constructs}

The $20$ bp DNA sequence, Eq.~(\ref{eq:Seq1}), used in MD simulations was
extracted from the $94$ bp E6-94 DNA sequence used in the previous DNA
cyclization experiment \cite{Cloutier:2004fg},
\begin{equation}
\begin{array}{l}
5'-\texttt{GTGCGCACGAAATGCTATGC}-3'\\
3'-\texttt{CACGCGTGCTTTACGATACG}-5'
\end{array}.
\label{eq:Seq1}
\end{equation}
The basepairs are indexed by $i$, in the $5'$ to $3'$ direction of the top
strand (also referred to as ``Strand \Rmnum{1}'') of the dsDNA segment.
Smoothly bent \textit{B}-form DNA were generated by the program \texttt{X3DNA}
\cite{Lu:2003ug} and served as the initial conformations for the simulations
(Fig.~S1 in the Supporting Material). A nick on nick-containing DNA of the
same sequence was generated by deleting the phosphate group on one strand
between two adjacent basepairs straddling the nicked site, thus leaving the two
broken backbone ends hydrolyzed (Fig.~S2).

\subsection*{Spring constraints}

Contractile springs with various equilibrium lengths/spring constants are
connected between the two nitrogenous bases of the $2^{\text{nd}}$ basepair and
those of the $19^{\text{th}}$ basepair to induce bending of different levels.
Force is distributed among their base atoms according to atomic weights. A
particular spring constraint is denoted by $\left\{\kappa; l\right\}$, where
$\kappa$ is the spring constant in units of pN/nm and $l$ is the equilibrium
length of the spring in units of nm.

Two different types of simulations were performed with two different purposes.
One set of simulations produced a sharply bent DNA to examine defect excitation
and test if the defect causes the sharp DNA bending. For this purpose, we used
springs of zero equilibrium length, adjusting their spring constants to
generate forces greater than the buckling transition force to bend the DNA, yet
small enough to provide sufficient time to observe both defect excitation and
the development of DNA bending.

The other set of simulations scanned the free energy landscape of DNA before
and after defect excitation based on umbrella sampling. Springs with finite
equilibrium lengths were used to constrain the end-to-end distance fluctuations
near a series of targeted values. The spring constant was determined to be
sufficiently stiff to constrain the regional fluctuations, yet soft enough to
allow overlapping of regional fluctuations that is needed for umbrella
sampling. Because of the need to constrain the narrow regional fluctuations,
these simulations are much stiffer than the first set of simulations.

\subsection*{MD simulations}

The DNA was placed in $150$ mM NaCl solution using explicit TIP3P water model
\cite{Jorgensen:1983fl} (see Supporting Methods: Unit cell preparation in the
Supporting Material). The MD simulations were then performed using
\texttt{GROMACS} version 4.5.5 \cite{Hess:2008db, vanderSpoel:2005hz,
Pronk:2013ef} under recent Parm99 force field with ParmBSC0 corrections
\cite{CheathamIII:1999ek, Perez:2007ek}. MD simulations are usually $70$ ns
each consisting of $50$ ns equilibration stage and $20$ ns production stage.
These simulations were executed using periodic boundary conditions under NVT
ensemble, with a constant volume of $\sim 1170$ nm$^3$ and a constant
temperature of $300$ K (or $290$, $310$ K with investigations into the effects
of temperature). The conformational representatives during the production stage
were used for extracting interested ensemble averages, such as the averages of
end-to-end distances, $\left< d \right>$. Before any constrained simulations,
an unconstrained simulation of $20$ bp DNA was conducted for $70$ ns as control
during which DNA maintained a regular helical structure with expected helical
repeat and pitch (Fig.~S3).

Macroscopic configuration information of DNA was extrapolated using local
basepair coordinates with the $x$ and $y$ directions in the basepair plane and
the $z$ direction perpendicular to the basepair plane (see Fig.~S4 and
Supporting Methods: Basepair coordinates in the Supporting Material). For
example, the bending angle between $i^\text{th}$ and $\left( i+\Delta
\right)^\text{th}$ basepairs, defined by $\theta_{i,i+\Delta} = \cos^{-1}
\left( \hat{z}_i \cdot \hat{z}_{i+\Delta} \right)$, where $i = 2, 3, \cdots,
19-\Delta$, can be calculated for any instantaneous conformation of DNA.

\section{RESULTS}

\subsection*{DNA bending responses under weak and strong spring constraints}

At a temperature of $300$ K, a $20$ bp DNA segment was forced to bend
connecting to the $2^{\text{nd}}$ and $19^{\text{th}}$ basepairs of the DNA
with a spring of zero equilibrium length (i.e., $\left\{\kappa; 0\right\}$; see
Fig.~S1 for initial DNA structure). Therefore, the region of DNA subject to the
spring constraint has $18$ basepairs and $17$ basepair steps. A total of $280$
DNA conformations were obtained in $14$ independent simulations under various
spring constraints in the range of $\kappa \in (8.0, 85.0)$ pN/nm from $50$ to
$70$ ns at regular $1$ ns intervals (Fig.~\ref{fig:2Overview}). During each
simulation, the constrained distance $d_{\left\{\kappa; 0\right\}}$ between the
center-of-mass of the atom groups in the two connected bases was monitored. In
addition, within each DNA basepairs the inter-distances of atoms involved in
hydrogen-bond formation, $h_{i,j}$ ($i$ denotes the basepair index and $j$
denotes the $j^{\text{th}}$ hydrogen bond in that basepair), were also
monitored.

Two representative snapshots of conformations at $t=60$ ns during simulations
confined by a weaker spring ($\kappa = 16.6$ pN/nm) and a stronger spring
($\kappa = 28.2$ pN/nm) reveal completely different bending responses
(Figs.~\ref{fig:3Compair}\textbf{A} and~\ref{fig:3Compair}\textbf{C}). The DNA
under the constraint of the stronger spring assumes a much more severely bent
conformation than DNA under the weaker spring, which contains disrupted
basepairs highlighted with the red shadowed area. The backbones of the $280$
DNA conformations can be classified into two distinctive groups based on the
level of bending (Fig.~\ref{fig:2Overview}, obtained from $14$ independent
simulations conducted with a wide range of spring constraints). In the weakly
bent group obtained at $\kappa < 20.0$ pN/nm, the end-to-end distances of DNA
are longer than that of the initial DNA (red line), indicating a balance
between the spring elastic energy and the DNA bending energy, which relaxed DNA
to a more straight conformation. In the sharply bent group obtained at $\kappa
> 25.0$ pN/nm, the end-to-end distances are significantly shorter than that of
the initial DNA. This indicates that the stiff springs out-competed the DNA
bending elasticity and forced DNA to collapse utile the two ends collided into
each other, which was accompanied with disruptions of DNA basepairs (e.g., the
shadowed region in Fig.~\ref{fig:3Compair}\textbf{C}).

We investigated the weakly bent DNA under $\kappa = 16.6$ pN/nm for its
structural details. The final value of $\left< d_{\left\{\kappa; 0\right\}}
\right>$, which was averaged over the last $20$ ns data out of $70$ ns
simulation, was $\sim 4.65$ nm. This is slightly longer than the initial value
$d_\text{ini} \approx 4.20$ nm indicating the tendency of DNA to relax to a
more straight conformation. However, $\left< d_{\left\{\kappa; 0\right\}}
\right>$ is still slightly shorter than the expected contour length of
\textit{B}-DNA of $17$ basepair steps ($\sim 5.43$ nm), indicating a weakly
bent conformation due to this spring constraint. The minimal and maximal
lengths of hydrogen bonds in each weakly bent basepair, which were averaged in
the last $20$ ns, $\left< \min (h_{i,j}) \right>$ and $\left< \max \left(
h_{i,j} \right) \right>$ completely overlap with those of control ($\kappa=0$
pN/nm). This indicates that the weakly bent DNA remained intact throughout $70$
ns simulation (Fig.~\ref{fig:3Compair}\textbf{B}). The hydrogen-bond length
fluctuates within $0.26 - 0.33$ nm with an average value $\sim 0.30$ nm, which
is consistent with hydrogen-bond lengths in the crystal structures of
\textit{B}-form DNA \cite{FonsecaGuerra:1999ki}. Thus, hereafter a basepair is
considered as Watson-Crick basepair when all its hydrogen-bond lengths are $<
0.33$ nm.

On the other hand, the \textit{B}-DNA became unstable when $\kappa > 25.0$
pN/nm, resulting in sharply bent DNA conformations with very short final
$\left< d_{\left\{\kappa; 0\right\}} \right> < 2.30$ nm (Fig.~S5). Considering
volume exclusion, this suggests that only a distance of DNA diameter separates
the two DNA ends. Such sharp DNA bending is always accompanied with disruption
of DNA basepairs. As an example, the conformation snapshot at $60$ ns of a
simulation with $\kappa = 28.2$ pN/nm contains a localized sharp bend near the
middle of the DNA (Fig.~\ref{fig:3Compair}\textbf{C}). The hydrogen-bonding
profile, $\left< \min, \max \left( h_{i,j} \right) \right>$, of this sharply
bent DNA (Fig.~\ref{fig:3Compair}\textbf{D}) clearly indicates that the
$11^{\text{th}} - 13^{\text{th}}$ basepairs are disrupted.

\subsection*{Basepair disruption results in localized sharp DNA bending}

We then sought to analyze the influence of local DNA basepair disruption in
sharply bent DNA on the overall shape of DNA. Thus, we calculated the bending
angle between the intact $10^{\text{th}}$ and $14^{\text{th}}$ basepairs that
straddles the disrupted region of DNA bent under $\kappa=28.2$ pN/nm using
$\theta_{10,14} = \cos^{-1} \left(\hat{z}_{10} \cdot \hat{z}_{14}\right)$,
where $\hat{z}_i$ describes the direction perpendicular to the $i^{\text{th}}$
basepair plane (see Materials and Methods and Fig.~S4 for details). The first
row in Fig.~\ref{fig:4Dynamics} shows that evolution of $\theta_{10,14}$ from
initial $\sim 40^{\circ}$ toward larger bending angle began immediately after
the simulation started. Saturated local bending was reached within $10$ ns, and
remained at a high bending level at $\sim 160^{\circ}$ throughout the remainder
of the simulation.

We also plotted the evolution of bending angles of two unaffected regions of
the same length ($\theta_{6,10}$ and $\theta_{14,18}$, \textit{row~1} of
Fig.~\ref{fig:4Dynamics}). Synchronized with DNA kink formation of
$\theta_{10,14}$, these bending angles relaxed from initial $\sim 40^{\circ}$
to values of $\sim 30^{\circ}$ and $\sim 10^{\circ}$ within $10$ ns,
respectively, and remained at these low bending levels throughout the remainder
of the simulation. These results indicate the kink formation relaxes the rest
of the DNA to a more straight conformation.

We further examined the correlation between the localized kink formation and
the disruption of basepairs. Time traces of $h_{i,j}$ for the three affected
\texttt{A$=$T} basepairs $i=11,12,13$ are shown in rows~2-4 of
Fig.~\ref{fig:4Dynamics}. These results reveal that the $11^{\text{th}}$
basepair remained intact in the first $\sim 48$ ns, and was then disrupted
between $\sim 48$ and $56$ ns, after which it fluctuated between disrupted and
intact states. The $12^{\text{th}}$ and $13^{\text{th}}$ basepairs opened up
within $10$ ns and remained disrupted. Clearly, DNA kink formation and
disruptions of these basepairs are highly correlated. Hence, we conclude that
basepair disruption causes kink development. We also note that sharply bent DNA
containing disrupted basepairs could be restored into a straight
\textit{B}-form DNA conformation within dozens of nanoseconds upon removal of
the spring constraint from the DNA (Fig.~S6).

\subsection*{Central localization of defects}

The development of similar localized kinks was observed in all twelve
independent simulations using $\kappa > 25.0$ pN/nm, which was accompanied with
basepair disruptions at kinked locations. These kinks primarily located around
the same region near the center, are likely due to the high curvature at the
center under our bending geometry.

Fig.~\ref{fig:5DefectLocation}\textbf{A} plots the hydrogen-bonding profiles,
$\left< \min, \max \left( h_{i,j} \right) \right>$ against $i$ values averaged
over the last $20$ ns (from all twelve independent simulations with $\kappa >
25.0$ pN/nm). This plot reveals that the disrupted basepairs occur around the
same region near DNA center that are \texttt{AT}-rich (i.e.,
$5'-\texttt{AAAT}-3'$, the $10^{\text{th}} - 13^{\text{th}}$ basepairs). One
possible cause for the central localization of basepair disruption is that the
largest curvature occurs at the center (Fig.~S7). Alternatively, it may be due
to the less stable noncovalent interactions of \texttt{AT}-rich region in the
middle of our DNA. Based on the unified NN basepair parameters by SantaLucia
\cite{SantaLucia:1998uz}, melting \texttt{A$=$T} next to \texttt{A$=$T}
basepairs is more feasible energetically than melting \texttt{A$=$T} next to
\texttt{G$\equiv$C} or melting \texttt{G$\equiv$C} next to \texttt{A$=$T}
basepairs, and melting \texttt{G$\equiv$C} next to \texttt{G$\equiv$C}
basepairs is the hardest.

To see which factor predominates in central localization, we shifted the entire
sequence tail-to-head by $2$ bp and replaced the central \texttt{AT}-rich
island at the $10^{\text{th}} - 13^{\text{th}}$ basepairs with
$5'-\texttt{CGAA}-3'$. Five independent simulations under different level of
strong bending using $\left\{\kappa; 0\right\}$ spring constraints with $\kappa
> 25.0$ pN/nm were conducted for $70$ ns. The overlay of the resulting
hydrogen-bonding profiles in Fig.~\ref{fig:5DefectLocation}\textbf{B} shows
that basepair disruptions still occurred at the central region, mainly at the
$10^{\text{th}} -11^{\text{th}}$ basepairs (i.e., \texttt{G$\equiv$C}
basepairing), and $12^{\text{th}}$ basepairs (i.e., \texttt{A$=$T}
basepairing). Taken together, these results suggest that the central
localization of the basepair disruptions is mainly caused by the high curvature
at the center of DNA, while the sequence effects are minimal under our bending
constraints.

\subsection*{DNA conformational free energy and force distance curves}

To understand the mechanics of DNA under bending, we calculated the DNA
conformational free energy as a function of end-to-end distance, ${\cal A}
\!\left(d \right)$, as well as the force required to maintain an end-to-end
distance, $f \! \left( d\right)$, using umbrella sampling for DNA under twelve
different spring constraints $\left\{248.9; l_m\right\}$ indexed by $m$. Here,
a fixed stiff spring constant of $\kappa = 248.9$ pN/nm was used in all
simulations to ensure that the end-to-end distance of DNA fluctuates near the
equilibrium spring length of $l_m$. A series of $l_m$ values ($5.27, 5.18,
4.94, 4.79, 4.56, 4.31, 4.17, 4.16, 3.80, 3.37, 3.01$, and $2.85$ nm) where
$l_1>l_2>\cdots>l_{12}$ were used to produce different levels of bending
constraint. The global unbiased ${\cal A} \!\left(d \right)$ was then obtained
based on these constrained local fluctuations using the standard weighted
histogram analysis method \texttt{g\_wham} \cite{Kumar:1992bv, Hub:2010ex} (see
details in Supporting Methods: Umbrella sampling in the Supporting Material).

The twelve constrained simulations led to nine segments with intact DNA
basepairs ($m=1, 2, \cdots, 9$) and three segments containing disrupted
basepairs in the region of $11^{\text{th}} - 13^{\text{th}}$ basepairs ($m=10,
11, 12$) in the last $20$ ns of total $50$ ns simulations. The inset of
Fig.~\ref{fig:6Profile} shows ${\cal A} \!\left(d \right)$ of \textit{B}-form
DNA obtained from nine intact DNA simulations (\textit{dark-red solid line}),
which contains a single energy minimum (set as $0$ $k_\text{B}T$) at
$d_\text{e} \approx 5.43$ nm. A DNA rise of $\sim 0.32$ nm/bp estimated by
$d_\text{e}/17$ is consistent with expected DNA rise of $0.33 \pm 0.02$ nm/bp
in the \textit{B}-form DNA duplex \cite{Olson:2001cf}. Note that there are $17$
basepair steps between the two spring-connected basepairs. We also obtained the
${\cal A} \!\left(d \right)$ for defect-containing DNA (\textit{dark-red dotted
line}, obtained with three simulations of DNA with disrupted basepairs), which
appears to have a smaller slope than the ${\cal A} \!\left(d \right)$ of
\textit{B}-form DNA. Because the umbrella sampling analysis was performed
separately for the each type of DNA, the ${\cal A} \!\left(d \right)$ profiles
have an undetermined offset from each other. Upon shifting the ${\cal A}
\!\left(d \right)$ of defect-containing DNA to match that of \textit{B}-form
DNA at their overlapping region, we noted that this shift does not affect the
calculation of $f \! \left( d\right)$.

A continuous force-distance curve could be obtained by $f \! \left( d\right)=-
\partial {\cal A}\!\left(d \right) / \partial d$. The $f \! \left( d\right)$ of
\textit{B}-form DNA is shown in Fig.~\ref{fig:6Profile} (\textit{dark-red solid
line}). This curve overlaps with results obtained by a direct readout through
$f\left(\left<d_{\left\{\kappa; l_m \right\}}\right>\right) = \kappa \times
\left(\left<d_{\left\{\kappa; l_m \right\}}\right> -l_m\right)$, where
$\left<d_{\left\{\kappa; l_m \right\}}\right>$ is the average end-to-end
distance under a particular spring constraint $\left\{248.9; l_m \right\}$
(corresponding \textit{dark-red dots}). As expected, at the equilibrium
distance $d_\text{e} \approx 5.43$ nm, the $f \! \left(d_\text{e} \right)=0$
pN. When $d$ is slightly shorter than $d_\text{e}$, the $f \! \left( d\right)$
increases linearly as $d$ decreases. The axial Young's modulus of DNA is
estimated to be $Y=\left( {\Delta f} / {\Delta d} \right) \left( L / S\right)
\approx 300$ $\text{pN/nm}^2$ as a result of this linear stress-strain relation
(with the contour length $L \approx d_\text{e}$, cross section $S=\pi R^2$, and
radius $R=1.0$ nm). The bending persistence length is estimated to be $A=\beta
Y I \approx 57.0$ nm, which is close to $53.4 \pm 2.3$ nm previous determined
in single-DNA stretching experiments \cite{Bustamante:1994wp}.

A transition from the initial linear force-distance curve ($d > 4.80$ nm) to a
nearly flatten profile ($4.00 < d < 4.60$ nm) occurs during decreasing $d$ in
conditions where $4.80 > d > 4.60$ nm, which corresponds to a force range of
$70 - 85$ pN. This behavior can be explained by classical Euler buckling
instability of elastic rods. Here, $f_\text{c}=\beta^{-1} \pi^2 A / {L^2}$
predicts a critical force at the onset of the rod bending (i.e., buckling
transition), when $L \ll A$, where $A$ is bending persistence length, and $L$
is DNA contour length. Using the simulated $A \approx 57.0$ nm, the
$f_\text{c}$ value is estimated to be $79.1$ pN, which is in agreement with the
simulated force range. Thus, we have successfully predicted the Young's modulus
and the buckling transition force of \textit{B}-form DNA, which indicates that
the force field is suitable for simulating large scale of DNA mechanical
properties. The result also indicates that the overall shape of DNA has reached
equilibrium over a wide range of bending constraints within our simulation
time.

Similar simulations constrained by $\left\{248.9; l_m\right\}$ were also
performed for defect-containing DNA. The $f \! \left(d \right)$ obtained by $-
\partial {\cal A}\!\left(d \right) / \partial d$ (Fig.~\ref{fig:6Profile},
\textit{dark-red dotted line}) as well as with a direct readout (corresponding
\textit{dark-red dots}) are also in agreement with each other. These results
reveal a significantly decreased $f \! \left(d \right)$ by $\sim 50$ pN
compared to \textit{B}-DNA force plateau after the buckling transition,
indicating that the defect-containing DNA is more flexible. In comparison to
\textit{B}-DNA, $f \! \left(d \right)$ obtained for the defect-containing DNA
has a more rugged profile. This is because the defect-containing DNA does not
have well-defined structures due to different types of defects and varying
levels of transient stacking with nearby basepairs.

\subsection*{Effects of nick on the micromechanics of sharply bent DNA}

To obtain insights into the experimental mechanical anomaly of sharply bent DNA
that contained nicks, we investigated the effects of nick on the micromechanics
of sharply bent DNA. We first performed MD simulations constrained by a
zero-length spring with $\kappa = 28.2$ pN/nm (i.e., spring constraint of
$\left\{28.2; 0\right\}$) to generate sharply bent conformations for four DNA
segments containing a single nick at different locations along the top strand
(Fig.~S2, nicks between the $6^{\text{th}}$ and $7^{\text{th}}$, between the
$8^{\text{th}}$ and $9^{\text{th}}$, between the $11^{\text{th}}$ and
$12^{\text{th}}$, and between the $13^{\text{th}}$ and $14^{\text{th}}$
basepairs, explicitly). During simulations, the interbase distances between
the adjacent $\text{C4'}$ atoms along the sugar-phosphate backbone of the
nicked strand, $\delta_{i,i+1}$, were monitored. Here $i$ is the basepair
index, which indicates the numbering of $\text{C4'}$ atoms starting from the
$1^\text{st}$ basepair.

For each of the four nicked DNAs, sharp bending led to significantly increased
$\delta_{i,i+1}$ that straddles the nick, indicating separation of the two
nick-straddling $\text{C4'}$ atoms and their associated bases
(Fig.~\ref{fig:7NickDefect}). The separation of the two $\text{C4'}$ atoms is
either caused by strand separation involving a few melted basepairs near the
nick (hereafter referred to as ``peeled'') or by unstacked basepairs straddling
the nick without hydrogen-bond disruptions (hereafter referred to as
``unstacked'') (Figs.~S8 and~S9). The selection between the two types of
defects depends on the two nick-straddling basepairs, where \texttt{G$\equiv$C}
basepairs are prone to unstacked defects and \texttt{A$=$T} basepairs are prone
to peeled defects (Fig.~S10).

Further analysis shows that the separation of the two $\text{C4'}$ atoms
straddling the nick is accompanied with a large bending angle developed at the
nicked position, which in turn relaxes the rest of DNA into a less bent
\textit{B}-form conformation. An example of this basepair separation is shown
in Fig.~\ref{fig:8NickDyn}\textbf{A}, where the nick is located between the
$8^{\text{th}}$ and $9^{\text{th}}$ basepairs. In the sharply bent
conformation, the $8^{\text{th}}$ and $9^{\text{th}}$ basepairs were unstacked,
leading to the increased $\delta_{8,9}$. The bending angle between the
$7^{\text{th}}$ and $10^{\text{th}}$ basepairs, $\theta_{7,10}$, rapidly
increased from the initial value of $\sim 30^\circ$ to $\sim 150^\circ$ within
$2$ ns after simulation began, and synchronized with the increase in
$\delta_{8,9}$. It also synchronized with relaxations of the three-basepair
step bending angles in the rest of DNA to more straight conformations, as shown
by the evolution of $\theta_{4,7}$ and $\theta_{10,13}$. In another example, a
similar nick between the $11^{\text{th}}$ and $12^{\text{th}}$ basepairs
promoted local sharp bending in the case of strand separation around the nick
(i.e., peeling) (Fig.~\ref{fig:8NickDyn}\textbf{B}). This peeling was caused by
disruptions of hydrogen bonds in the adjacent $11^{\text{th}}$,
$10^{\text{th}}$, and $9^{\text{th}}$ basepairs. The development of a large
bending angle around the nicked position synchronized with the relaxation of
the rest of DNA to a less bent \textit{B}-form conformation as well.

Then, using $\left\{248.9; l_m \right\}$-constrained simulations with umbrella
sampling analysis similar to those used with nick-free DNA, we obtained the
free energy-distance (${\cal A} \! \left( d\right)$) and force-distance ($f \!
\left( d\right)$) profiles for DNA containing a nick between the $11^\text{th}$
and $12^\text{th}$ basepairs (Fig.~\ref{fig:6Profile}, \textit{light-blue
lines}). Both profiles overlap with the intact nick-free DNA under weak bending
conditions, suggesting that the nicked DNA assumes \textit{B}-form at the
nicked sites and has similar bending elasticity to nick-free DNA under weak
bending conditions. However, increased bending leads to deviation of the
profiles from the \textit{B}-form profiles due to unstacking of the
$11^\text{th}$ and $12^\text{th}$ basepairs, which occurs between $4.00$ and
$5.20$ nm. Further bending ($d < 4.00$ nm) causes the peeling of $1-3$ bp of
nearby basepairs. The unstacking and peeling occurring at $d < 5.20$ nm results
in a force plateau of $< 40$ pN, which is significantly smaller than the
buckling transition force of \textit{B}-form DNA ($\sim 80$ pN). After the
flexible defect was excited at the nicked site, the $f \! \left(d \right)$
becomes rugged, which is similar to the profile observed for nick-free DNA with
basepair disruptions excited inside. Overall, these results demonstrate a
nick-dependent DNA softening through nick-promoted excitations of flexible
defects.

\subsection*{Effects of direction of bending on defect excitation}

To understand whether the direction of bending could affect the defect
excitation, we performed a series of $70$ ns simulations using zero-length
springs with a variety of spring constants (i.e., $\left\{\kappa; 0 \right\}$)
for both nick-free and nicked DNA bent into three evenly separated directions
(Fig.~\ref{fig:9DBIni}, \textit{top view}) denoted by \rmnum{1}, \rmnum{2}, and
\rmnum{3}. Each initial DNA conformation has a uniform bending angle per
basepair step of $\theta = 3.8^\circ$ by adjusting the tilt and roll angles of
the basepairs (see values in Table~S1 in the Supporting Material).

In simulations with nicked DNA, a single nick was introduced in the top strand
after the $11^\text{th}$ basepair. As shown in the side view of
Fig.~\ref{fig:9DBIni}, a local polar coordinate is defined at the nicked site
with the opposite-normal direction as the polar axes. In the local polar
coordinate, the angular positions of the nick are $+60^\circ$, $+180^\circ$,
and $-60^\circ$ in the DNAs bent into the directions \rmnum{1}, \rmnum{2}, and
\rmnum{3}, respectively. In the cases of $\pm 60^\circ$ nick positions (i.e.,
the bending directions \rmnum{1}~and \rmnum{3}), the nick is under a tensile
stress; for the $+180^\circ$ nick position (i.e., the bending direction
\rmnum{2}), the nick is under a compressive stress.

Simulations for the nick-free DNA were conducted under two spring constraints
of $\kappa=16.6$ and $28.2$ pN/nm. Under $\kappa=16.6$ pN/nm, the
\textit{B}-form DNA conformations remained intact throughout the simulations,
as demonstrated by the hydrogen-bonding profiles averaged from the last $20$ ns
simulations (Fig.~\ref{fig:10Isotropicity}\textbf{A}, \textit{top}). In
contrast, under the stronger constraint of $\kappa=28.2$ pN/nm, defect
excitation occurred near the middle of the DNAs regardless of direction of
bending (Fig.~\ref{fig:10Isotropicity}\textbf{A}, \textit{bottom}). These
results suggest that for nick-free DNA, the defect excitation is not sensitive
to direction of bending.

Similar simulations were performed for the nicked DNA under three spring
constraints of $\kappa=8.3$, $16.6$, and $28.2$ pN/nm. Under $\kappa=8.3$
pN/nm, defect excitation was not observed in any bending direction according to
their interbase distance profiles averaged in $50 - 70$ ns
(Fig.~\ref{fig:10Isotropicity}\textbf{B}, \textit{top}). However, under
$\kappa=16.6$ pN/nm, defect excitation only occurred in the bending direction
\rmnum{1}, which was located at the nicked site
(Fig.~\ref{fig:10Isotropicity}\textbf{B}, \textit{middle}). Considering that
under the same spring constant, defects cannot be excited for nick-free DNA in
any bending direction, this result is consistent with our conclusion that nicks
can facilitate defect excitation. In addition, because the defect excitation
only occurred in one bending direction within our simulation timescale, this
suggests that bending-induced nick-dependent defect excitation may have an
anisotropic dependence on the direction of bending. Under the strongest
constraint of $\kappa=28.2$ pN/nm, defects were excited at the nick regardless
of direction of bending (Fig.~\ref{fig:10Isotropicity}\textbf{B},
\textit{bottom}).

Overall, these results again demonstrate central localized defect excitation in
sharply bent nick-free DNA, and defect excitation at nicked sites in sharply
bent nick-containing DNA. In addition, a much weaker initial bending ($\sim
3.8^\circ$ per basepair step) was used here compared to that in
Figs.~\ref{fig:2Overview},~\ref{fig:3Compair},~\ref{fig:4Dynamics},~\ref{fig:5DefectLocation},~\ref{fig:6Profile},~\ref{fig:7NickDefect},
and~\ref{fig:8NickDyn} ($\sim 10^\circ$ per basepair step), which further
suggests that the main results of our simulations do not depend on the level of
initial bending.

\subsection*{Effects of temperature on nick-dependent defect excitation}

Because DNA basepair stability is sensitive to temperature and several sharp
DNA bending experiments were performed with different temperatures, we
investigated the effects of temperature at $290$, $300$, and $310$ K on the
nick-dependent defect excitation. For this, we used a spring with an
equilibrium length of $4.20$ nm and a spring constant of $248.9$ pN/nm (i.e., a
spring constant of $\left\{248.9; 4.20 \right\}$) to bend the DNA. Four
simulations were run for $50$ ns at each temperature to obtain the defect
excitation statistics. As defects did not occur in the nick-free DNA at these
temperatures with this spring constraint (data not shown), we decided to probe
the nick-dependent defect excitation at different temperatures with
$\left\{248.9; 4.20 \right\}$. As the nick-dependent defect excitation is
likely anisotropic, we introduced three nicks located after the $8^\text{th}$
basepair on Strand \Rmnum{1}, the $10^\text{th}$ basepair on Strand \Rmnum{2},
and the $12^\text{th}$ basepair on Strand \Rmnum{1} (Fig.~S11\textbf{A}). Under
any bending direction, the three nicks are exposed to different bending
orientations, which minimize the potential anisotropic effect. During
simulations, the interbase distances along Strand \Rmnum{1} and \Rmnum{2} were
monitored. They are denoted by $\delta_{i,i+1}^\text{\Rmnum{1}}$ and
$\delta_{i,i+1}^\text{\Rmnum{2}}$, respectively.

Under such bending constraints at $290$ K, defect excitation occurred at the
nicks. However, the defect excited state was not the predominant form and a
transient defected nick rapidly restacked (Fig.~S11\textbf{B}, \textit{top},
obtained at $290$ K). Their interbase distance profiles, $\left<
\delta^\text{\Rmnum{1}, \Rmnum{2}}_{i,i+1} \right>$, for both strands are
consistently similar to that of nick-free DNA (Fig.~\ref{fig:11TempC4},
\textit{top}), further indicating that the nicked sites predominantly exist in
the stacked \textit{B}-form conformation. The main mechanical effect of this
transient defect excitation is that the force in the spring to maintain such
bending constraint is $\sim 10\%$ lower than that for nick-free control DNA
(Table~1, for all four simulations at $290$ K averaged in the last $20$ ns).

\begin{table}[!htb]
\centering
\def\arraystretch{0.8}
\setlength\tabcolsep{0.4 cm}
\begin{tabular}{l c c c }
\toprule
 & \multicolumn{3}{c}{Force (pN)} \\
\cline{2-4}
 & $290$ K & $300$ K & $310$ K \\
\cline{1-4}
Run 1 & 69.9 &  1.5 & 35.6 \\
Run 2 & 69.7 & 29.3 & 27.2 \\
Run 3 & 67.7 & 12.5 & 16.0 \\
Run 4 & 66.4 & 19.9 & 20.0 \\
\cline{1-4}
Control & 81.7 & 83.3 & 82.5 \\
\botrule
\end{tabular}
\caption{Force ($\left<\kappa \times \left(d_{\left\{248.9;4.20 \right\}} - l
\right) \right>$) under the spring constraint of $\left\{248.9; 4.20 \right\}$
at different temperatures. The mean values of force in the spring (i.e., in
units of picoNewtons) are calculated in the last $20$ of $50$ ns simulations
for nicked DNA at three indicated temperatures, with four simulations performed
at each temperature denoted by runs $1-4$. For comparison, forces obtained on
nick-free DNA as control are $\sim 82$ pN even at $310$ K.}
\label{tab:TempEffect}
\end{table}

In sharp contrast, defect excited states dominated in all simulations performed
at both $300$ and $310$ K (see Fig.~S11\textbf{B}, \textit{bottom}, obtained at
$300$ K). The interbase distance profiles significantly deviate from the
\textit{B}-form behavior at one or more nicked sites (Fig.~\ref{fig:11TempC4},
\textit{middle} and \textit{bottom}). Furthermore, the force required to
maintain the same bending constraint is drastically reduced compared to that
for nick-free DNA, and that for nicked DNA at $290$ K (Table~1). Together,
these results indicate that the nick-dependent flexible defect excitation is
sensitive to temperature --- decreasing temperature can significantly inhibit
defect excitation at nicked sites.

\section{DISCUSSION}

In this work, we observed excitation of flexible DNA defects in sharply bent
DNA with disrupted basepairs. However, when the DNA contained a nick,
excitation of flexible defects predominantly occurred at the nicked site. Such
preferential excitation of flexible defects at nicked sites subsequently
absorbed bending to nicks and relaxed the level of bending elsewhere in the
DNA, which in turn suppressed defect excitation in nick-free region. These
results suggest that a nick in a DNA is a structurally weaker point compared to
the nick-free DNA region, which undergoes unstacking/peeling upon sharp
bending. This is in agreement with results obtained in a recent coarse-grained
MD simulation by Harrison et al. \cite{2015arXiv150609005H,
2015arXiv150609008H}. The idea that a nick is a weaker structural point was
also suggested by an earlier experiment showing that the unstacking/peeling
transition occurred preferentially at the nicked site with increasing
temperatures \cite{Protozanova:2004bo, Yakovchuk:2006bm}.

Previous $j$-factors measured for large DNA ($> 200$ bp) are consistent with
those predicted by the WLC model indicating that weakly bent DNA in large loops
could maintain a \textit{B}-form conformation at the hybridized double-nicked
region, and therefore satisfy the $\Omega$-boundary condition. This is
consistent with our results showing that under weak bending, a nick remains in
the stacked state with a \textit{B}-form conformation and bending stiffness.

The $j$-factor measurements strongly deviated from the canonical
WLC predictions when performed for shorter DNA fragments of $94-116$ bp. While
the $j$-factor was only slightly above the WLC prediction for $116$ bp
fragments, $j$-factors could be several orders of magnitude greater than WLC
predictions with shorter fragments of DNA \cite{Cloutier:2004fg,
Cloutier:2005et,Vafabakhsh:2012dt, Forties:2009kva}. The mechanics of the
unexpectedly high DNA looping probability was previously explained by
excitation of flexible defects inside DNA \cite{Cloutier:2004fg,
Cloutier:2005et, Yan:2004vt, Yan:2005vk, Wiggins:2005uu, Vafabakhsh:2012dt,
Le:2014iw}. Our results of the nick-dependent defect excitation in sharply bent
DNA provide another highly possible explanation: unstacking/peeling excitations
at the nick under increased level of bending implies violation of the
$\Omega$-boundary condition in looping experiment with short DNA fragments. As
shown with previous theoretical predictions \cite{Shimada:1984gw, Yan:2005vk,
Chen:2008ed}, if the two ends of the same DNA can meet in a kinked
conformation, the looping probability density is much higher compared to that
under the $\Omega$-boundary condition. Therefore, comparison between the
experimental $j$-factor measurements and theoretical predictions based on the
WLC model under the $\Omega$-boundary condition will lead to significantly
overestimated DNA bending flexibility.

Here, we discuss the possibilities of violating the $\Omega$-boundary condition
in the smFRET and the ligase-based $j$-factor measurements. In the smFRET
measurements, DNA looping is purely dependent on hybridization of the
complementary ends. Therefore, both nicks are under bending stress and can be
unstacked/peeled. The ligase-based $j$-factor measurements are more complex as
the looped DNA is covalently sealed by a subsequent ligation reaction for
quantification. An important question is whether the ligase enzyme only
recognizes a subset of the looped DNA, thereby imposing an additional
constraint on the conformation of the nicked sites. If the ligase can recognize
a kinked nick and use the binding energy to deform the nick into a conformation
that allows ligation, then the $\Omega$-boundary condition can be violated due
to the nick-dependent defect excitation. Furthermore, if a ligase can only
recognize a stacked \textit{B}-form nick, the $\Omega$-boundary condition can
still be violated because when a ligase seals a stacked nick in a double-nicked
DNA loop, the other nick can still remain in an unsealed unstacked state,
whereas the DNA loop is already irreversibly closed.

It is well known that the stacking energy between DNA basepairs has a
strong dependence on temperature \cite{SantaLucia:1998uz}, which may be related
to a discrepancy between two $j$-factor measurements for $94$ bp DNA fragments.
A canonical WLC elastic response of DNA was reported at $21^\circ$C
\cite{Du:2005gg}, which is in contrast to the mechanical anomaly observed at
$30^\circ$C \cite{Cloutier:2004fg,Cloutier:2005et}. Our simulations at
different temperatures revealed that the unstacking of the nick in a sharply
bent DNA is highly sensitive to temperature, which is significantly suppressed
when the temperature was reduced from $300$ to $290$ K. The observed trend of
temperature dependency of nick-dependent defect excitation in a sharply bent
DNA provides a possible explanation to the experimental discrepancy.

DNA mechanical anomaly was also reported by analyzing the elastic energy of
short dsDNA fragments, which were constrained in a sharply bent conformation
using a short ssDNA connecting the two dsDNA ends \cite{Qu:2010gz, Qu:2011jz}.
However, a preexisting nick was introduced to the middle of the dsDNA in those
experiments, while the interpretation of the intrinsic mechanical anomaly of
dsDNA relied upon the assumption that the nick remained in the \textit{B}-form
conformation in the experiments. According to our simulation, the apparent
anomaly observed in those experiments could also be explained by a
nick-dependent flexible defect excitation.

The mechanics of sharply bent DNA was also studied in sharply bent nick-free
DNA fragments. Shroff et al. \cite{Shroff:2005cw} bent a nick-free $25$ bp
($24$ basepair steps) dsDNA fragment using a $12$ nt ssDNA connecting the two
dsDNA ends. The work reported a tension in the ssDNA of $6 \pm 5$ pN, a few
times smaller than the buckling transition force ($\sim 30$ pN) expected from
the canonical WLC model, indicating mechanical anomaly in this sharply bent
DNA. As the level of bending in this experiment is much higher than that in
$\sim 100$ bp DNA minicircles (see Supporting Discussion in the Supporting
Material for details), it does not provide an answer to whether a similar
mechanical anomaly could occur in $\sim 100$ bp nick-free DNA minicircles.
Mechanical anomaly in severely sharply bent DNA can be explained by flexible
defect excitation inside DNA due to basepair disruption. It is consistent with
our simulations on nick-free DNA and an experiment reporting ssDNA formation in
covalently ligated $63-65$ bp DNA minicircles based on BAL-31 nuclease
digestion assay \cite{Du:2007be, Vologodskii:2013il}.

Deviation from the canonical WLC model was also reported based on analyzing the
bending angle distribution over short DNA contour length using atomic force
microscopy imaging in air. That study reported that $5-10$ nm DNA fragments
have a significantly higher probability for larger bending angle than that
predicted by the canonical WLC polymer model \cite{Wiggins:2006fu}. However,
one cannot exclude the possibility that perturbation during sample drying
processes might cause rare large DNA kinks. Indeed, this has been demonstrated
in a more recent atomic force microscopy imaging experiment carried out in
solution, which reported a normal bending angle distribution expected from the
canonical WLC polymer model for $\sim 10$ nm DNA fragments
\cite{PhysRevLett.112.068104}.

The micromechanics of DNA bending was also studied by analyzing the shapes of
$94$ bp DNA minicircles imaged using cryo-electron microscopy for three DNA
constructs: (\textit{i}) DNA contains two $2$ nt ssDNA gaps, (\textit{ii}) DNA
contains two nicks, and (\textit{iii}) DNA without either gap or nick
\cite{Demurtas:2009ju}. This study reported localized kinks formed in gapped
DNA only, indicating that flexible defects were not excited in either nicked or
nick-free DNA minicircles. However, as cryo-electron microscopy requires a
rapid (milliseconds) freezing step of the DNA samples, one cannot preclude the
possibility that an excited defect before cryo freezing could reanneal during
freezing process. Therefore, the results from this imaging study cannot be
directly compared with results from previous DNA looping experiments using
similar length of DNA.

Besides the aforementioned experimental efforts, mechanics of sharply bent DNA
was also investigated using full-atom MD simulations. Unstacked kinks were
observed to form in $94$ bp nick-free DNA minicircles at $300$ K using Parm94
force field \cite{Lankas:2006jz}. However, it has been known that
\textit{B}-DNA simulated using Parm94 have overpopulated $\alpha$/$\gamma$
transitions and geometric deviations from \textit{B}-DNA \cite{Perez:2007ek,
Perez:2008ft}; therefore, it is unclear whether the observed defect excitation
was caused by use of the Parm94 force field or it was an intrinsic elastic
response of DNA.

Is there any evidence supporting nick-independent flexible defect excitation in
$\sim 100$ bp DNA loops? To our knowledge, there are two pieces of evidence. A
recent full-atom MD simulation using Parm99 with ParmBSC0 correction reported
that deviation from the canonical WLC model occurred at bending angles $>
50^\circ$ with a short DNA fragment of $15$ bp ($14$ basepair steps). This
level of bending is comparable to that in a $94$ bp DNA loop in a planar
circular conformation (i.e., $14/94\times360^\circ \approx 54^\circ$);
therefore, this suggests that defects could potentially be excited inside DNA
under a similar level of bending constraints \cite{Curuksu:2009cv}. In
addition, a $j$-factor measurement by Forties et al. \cite{Forties:2009kva}
reported values slightly (less than fivefold) greater than the WLC prediction
under the $\Omega$-boundary condition on $116$ bp DNA at temperatures above
$30^\circ$C. The anomalous elasticity was observed for a DNA sequence
containing eight \texttt{TAT} repeats, which creates $16$ thermally weak
\texttt{AT} basepair steps \cite{SantaLucia:1998uz}, but not for another DNA of
the same length lacking such \texttt{TAT} repeats even at $37^\circ$C. As the
observed anomaly depends on the presence of multiple \texttt{TAT} repeats
inside DNA, their results cannot be explained by nick-dependent defect
excitation. However, the strong dependence on the presence of multiple
\texttt{TAT} repeats raises the question whether the same mechanism could
explain the observed mechanical anomaly in other DNA cyclization experiments,
as DNAs used in these experiments do not contain such specifically inserted
weak basepair repeats \cite{Cloutier:2004fg, Cloutier:2005et,Vafabakhsh:2012dt,
Le:2014iw}.

Taken together, our simulations suggest that when a looped short DNA contains
nicks, the nicks have the weakest mechanical stability and are prone to develop
flexible defects compared to other sites in the DNA. However, as defect
excitations at the nicks and in the nick-free DNA region are in thermodynamic
competition, which is a predominant factor is not trivial. This obviously
depends on the number of weak basepair steps in the nick-free DNA region. A
crudest estimate of the possibility $P$ of having at least one disrupted weak
basepair steps is: $P=1-(1-p)^N$, where $p$ is the probability of a particular
weak basepair step in the disrupted state and $N$ is the number of weak
basepair steps. As $P$ increases with $N$, at large $N$ values defect
excitation at such weak basepair steps may be able to outcompete that at the
nicks and becomes the dominant factor. Therefore, their competition likely
depends on many solution factors (such as temperature, salt, and pH that affect
DNA basepair stability), sequence composition, size of DNA (the shorter the
less $N$ of weak basepair steps), and the level of bending. In addition, for
looped DNA the level of twist has a significant effect on DNA basepair
stability \cite{Harris:2008hc, Liverpool:2008db, Mitchell:2013cp}. Considering
the importance of this level of DNA bending in $\sim 100$ bp loops, the
outstanding scientific controversy it has caused and the complex dependence on
the above-mentioned experimental conditions, new experiments using nick-free
DNA are warranted to readdress this important question by systematically
elucidating the roles of each of these contributing factors.

\section*{SUPPORTING MATERIAL}

Supporting Materials and Methods, Supporting Discussion, eleven figures and one
table are available at http://arxiv.org/.

\section*{AUTHOR CONTRIBUTIONS}

J.Y., P.D., J.R.C.v.d.M. conceived the study; P.C. and L.D. performed the MD
simulation; P.C., J.Y., and H.C. interpreted and analyzed the data; P.C. and
J.Y. wrote the article.

\section*{ACKNOWLEDGMENTS}

The authors are grateful to John Marko (Northwestern University) and Ralf
Bundschuh (Ohio State University) for valuable discussions.

The work is funded by the Mechanobiology Institute at the National University
of Singapore, by the Ministry of Education Singapore Academic Research Fund
Tier 2 (grant No. MOE2013-T2-1-154) and Tier 3 (grant No. MOE2012-T3-1-001) (to
J.Y.), and by the National Research Foundation Singapore through the
Singapore-MIT Alliance for Research and Technology's research program in
BioSystems and Micromechanics (to P.D.).

\section*{SUPPORTING CITATIONS}

References \cite{Yamakawa:1972jk, Lu:2008kb,
:/content/aip/journal/jcp/126/1/10.1063/1.2408420, Parrinello:1981it,
Clowney:1996kr, Horn:1987hf, Torrie:1977hs, Cocco:2004fo} appear in the
Supporting Material.



\begin{thebibliography}{66}
\providecommand{\url}[1]{\texttt{#1}}
\providecommand{\urlprefix}{ }

\bibitem[Richmond and Davey(2003)]{Richmond:2003hm}
Richmond, T.~J., and C.~A. Davey, 2003.
\newblock {The structure of DNA in the nucleosome core}.
\newblock \emph{Nature} 423:145--150.

\bibitem[Oehler et~al.(1994)Oehler, Amouyal, Kolkhof, von Wilcken-Bergmann, and
  Muller-Hill]{Oehler:1994vz}
Oehler, S., M.~Amouyal, P.~Kolkhof, B.~von Wilcken-Bergmann, and
  B.~Muller-Hill, 1994.
\newblock {Quality and position of the three lac operators of \textit{E. coli}
  define efficiency of repression.}
\newblock \emph{EMBO J.} 13:3348--3355.

\bibitem[Hagerman(1988)]{Hagerman:1988}
Hagerman, P.~J., 1988.
\newblock {Flexibility of DNA}.
\newblock \emph{Annu. Rev. Bioph. Biom.} 17:265--286.

\bibitem[Shore et~al.(1981)Shore, Langowski, and Baldwin]{Shore:1981wt}
Shore, D., J.~Langowski, and R.~L. Baldwin, 1981.
\newblock {DNA flexibility studied by covalent closure of short fragments into
  circles}.
\newblock \emph{Proc. Natl. Acad. Sci. USA.} 78:4833--4837.

\bibitem[Doi and Edwards(1986)]{Doi:1986ug}
Doi, M., and S.~F. Edwards, 1986.
\newblock {The Theory of Polymer Dynamics}.
\newblock Clarendon Press, Oxford, UK.

\bibitem[Smith et~al.(1992)Smith, Finzi, and Bustamante]{Smith:1992em}
Smith, S.~B., L.~Finzi, and C.~Bustamante, 1992.
\newblock {Direct mechanical measurements of the elasticity of single DNA
  molecules by using magnetic beads}.
\newblock \emph{Science} 258:1122--1126.

\bibitem[Marko and Siggia(1995)]{Marko:1995}
Marko, J.~F., and E.~D. Siggia, 1995.
\newblock {Stretching DNA}.
\newblock \emph{Macromolecules.} 28:8759--8770.

\bibitem[Cloutier and Widom(2004)]{Cloutier:2004fg}
Cloutier, T.~E., and J.~Widom, 2004.
\newblock {Spontaneous sharp bending of double-stranded DNA}.
\newblock \emph{Mol. Cell.} 14:355--362.

\bibitem[Vafabakhsh and Ha(2012)]{Vafabakhsh:2012dt}
Vafabakhsh, R., and T.~Ha, 2012.
\newblock {Extreme bendability of DNA less than 100 base pairs long revealed by
  single-molecule cyclization}.
\newblock \emph{Science} 337:1097--1101.

\bibitem[Luger et~al.(1997)Luger, M{\"a}der, Richmond, Sargent, and
  Richmond]{Luger:1997gf}
Luger, K., A.~W. M{\"a}der, R.~K. Richmond, D.~F. Sargent, and T.~J. Richmond,
  1997.
\newblock {Crystal structure of the nucleosome core particle at 2.8 {\AA}
  resolution}.
\newblock \emph{Nature} 389:251--260.

\bibitem[Davey et~al.(2002)Davey, Sargent, Luger, Maeder, and
  Richmond]{Davey:2002ir}
Davey, C.~A., D.~F. Sargent, K.~Luger, A.~W. Maeder, and T.~J. Richmond, 2002.
\newblock {Solvent mediated interactions in the structure of the nucleosome
  core particle at 1.9 {\AA} resolution}.
\newblock \emph{J. Mol. Biol.} 319:1097--1113.

\bibitem[Jacobson and Stockmayer(1950)]{Jacobson:1950kv}
Jacobson, H., and W.~H. Stockmayer, 1950.
\newblock {Intramolecular reaction in polycondensations. I. The theory of
  linear systems}.
\newblock \emph{J. Chem. Phys.} 18:1600.

\bibitem[Shore and Baldwin(1983)]{Shore:1983jm}
Shore, D., and R.~L. Baldwin, 1983.
\newblock {Energetics of DNA twisting: I. Relation between twist and
  cyclization probability}.
\newblock \emph{J. Mol. Biol.} 170:957--981.

\bibitem[Du et~al.(2005)Du, Smith, Shiffeldrim, Vologodskaia, and
  Vologodskii]{Du:2005gg}
Du, Q., C.~Smith, N.~Shiffeldrim, M.~Vologodskaia, and A.~Vologodskii, 2005.
\newblock {Cyclization of short DNA fragments and bending fluctuations of the
  double helix}.
\newblock \emph{Proc. Natl. Acad. Sci. USA.} 102:5397--5402.

\bibitem[{Claus Roll} et~al.(1998){Claus Roll}, {Christophe Ketterl\'{e}},
  {Val\'{e}rie Faibis}, Fazakerley, , and Boulard]{ClausRoll:1998dl}
{Claus Roll}, {Christophe Ketterl\'{e}}, {Val\'{e}rie Faibis}, G.~V.
  Fazakerley, , and Y.~Boulard, 1998.
\newblock {Conformations of nicked and gapped DNA structures by NMR and
  molecular dynamic simulations in water}.
\newblock \emph{Biochemistry} 37:4059--4070.

\bibitem[Hyz et~al.(2011)Hyz, Bocian, {Kaw\k{e}cki}, Bednarek, Sitkowski, and
  Kozerski]{Hyz:2011dv}
Hyz, K., W.~Bocian, R.~{Kaw\k{e}cki}, E.~Bednarek, J.~Sitkowski, and
  L.~Kozerski, 2011.
\newblock {A dumbbell double nicked duplex dodecamer DNA with a PEG6 tether}.
\newblock \emph{Org. Biomol. Chem.} 9:4481--4486.

\bibitem[Taylor and Hagerman(1990)]{Taylor:1990ve}
Taylor, W.~H., and P.~J. Hagerman, 1990.
\newblock {Application of the method of phage T4 DNA ligase-catalyzed
  ring-closure to the study of DNA structure: II. NaCl-dependence of DNA
  flexibility and helical repeat}.
\newblock \emph{J. Mol. Biol.} 212:363--376.

\bibitem[Cloutier and Widom(2005)]{Cloutier:2005et}
Cloutier, T.~E., and J.~Widom, 2005.
\newblock {DNA twisting flexibility and the formation of sharply looped
  protein-DNA complexes.}
\newblock \emph{Proc. Natl. Acad. Sci. USA.} 102:3645--3650.

\bibitem[Yan and Marko(2004)]{Yan:2004vt}
Yan, J., and J.~F. Marko, 2004.
\newblock {Localized single-stranded bubble mechanism for cyclization of short
  double helix DNA}.
\newblock \emph{Phys. Rev. Lett.} 93:108108.

\bibitem[Yan et~al.(2005)Yan, Kawamura, and Marko]{Yan:2005vk}
Yan, J., R.~Kawamura, and J.~F. Marko, 2005.
\newblock {Statistics of loop formation along double helix DNAs}.
\newblock \emph{Phys. Rev. E Stat. Nonlin. Soft Matter Phys.} 71:061905.

\bibitem[Wiggins et~al.(2005)Wiggins, Phillips, and Nelson]{Wiggins:2005uu}
Wiggins, P.~A., R.~Phillips, and P.~C. Nelson, 2005.
\newblock {Exact theory of kinkable elastic polymers}.
\newblock \emph{Phys. Rev. E Stat. Nonlin. Soft Matter Phys.} 71:021909.

\bibitem[Destainville et~al.(2009)Destainville, Manghi, and
  Palmeri]{Destainville:2009fw}
Destainville, N., M.~Manghi, and J.~Palmeri, 2009.
\newblock {Microscopic mechanism for experimentally observed anomalous
  elasticity of DNA in two dimensions}.
\newblock \emph{Biophys. J.} 96:4464--4469.

\bibitem[Le and Kim(2014)]{Le:2014iw}
Le, T.~T., and H.~D. Kim, 2014.
\newblock {Probing the elastic limit of DNA bending}.
\newblock \emph{Nucleic Acids Res.} 42:10786--10794.

\bibitem[Chen and Yan(2008)]{Chen:2008ed}
Chen, H., and J.~Yan, 2008.
\newblock {Effects of kink and flexible hinge defects on mechanical responses
  of short double-stranded DNA molecules}.
\newblock \emph{Phys. Rev. E Stat. Nonlin. Soft Matter Phys.} 77:041907.

\bibitem[Vologodskii et~al.(2013)Vologodskii, Du, and
  Frank-Kamenetskii]{Vologodskii:2013ca}
Vologodskii, A.~V., Q.~Du, and M.~D. Frank-Kamenetskii, 2013.
\newblock {Bending of short DNA helices}.
\newblock \emph{Artif. DNA PNA XNA} 4:1--3.

\bibitem[Lu and Olson(2003)]{Lu:2003ug}
Lu, X.-J., and W.~K. Olson, 2003.
\newblock {\texttt{3DNA}: a software package for the analysis, rebuilding and
  visualization of three-dimensional nucleic acid structures}.
\newblock \emph{Nucleic Acids Res.} 31:5108--5121.

\bibitem[Jorgensen et~al.(1983)Jorgensen, Chandrasekhar, Madura, Impey, and
  Klein]{Jorgensen:1983fl}
Jorgensen, W.~L., J.~Chandrasekhar, J.~D. Madura, R.~W. Impey, and M.~L. Klein,
  1983.
\newblock {Comparison of simple potential functions for simulating liquid
  water}.
\newblock \emph{J. Chem. Phys.} 79:926--935.

\bibitem[Hess et~al.(2008)Hess, Kutzner, van~der Spoel, and
  Lindahl]{Hess:2008db}
Hess, B., C.~Kutzner, D.~van~der Spoel, and E.~Lindahl, 2008.
\newblock {\texttt{GROMACS} 4: algorithms for highly efficient, load-balanced,
  and scalable molecular simulation}.
\newblock \emph{J. Chem. Theory Comput.} 4:435--447.

\bibitem[van~der Spoel et~al.(2005)van~der Spoel, Lindahl, Hess, Groenhof,
  Mark, and Berendsen]{vanderSpoel:2005hz}
van~der Spoel, D., E.~Lindahl, B.~Hess, G.~Groenhof, A.~E. Mark, and H.~J.~C.
  Berendsen, 2005.
\newblock {\texttt{GROMACS}: fast, flexible, and free}.
\newblock \emph{J. Comput. Chem.} 26:1701--1718.

\bibitem[Pronk et~al.(2013)Pronk, P{\'a}ll, Schulz, Larsson, Bjelkmar,
  Apostolov, Shirts, Smith, Kasson, van~der Spoel, Hess, and
  Lindahl]{Pronk:2013ef}
Pronk, S., S.~P{\'a}ll, R.~Schulz, P.~Larsson, P.~Bjelkmar, R.~Apostolov, M.~R.
  Shirts, J.~C. Smith, P.~M. Kasson, D.~van~der Spoel, B.~Hess, and E.~Lindahl,
  2013.
\newblock {\texttt{GROMACS} 4.5: a high-throughput and highly parallel open
  source molecular simulation toolkit}.
\newblock \emph{Bioinformatics.} 29:845--854.

\bibitem[Cheatham et~al.(1999)Cheatham, Cieplak, and
  Kollman]{CheathamIII:1999ek}
Cheatham, T.~E., 3rd, P.~Cieplak, and P.~A. Kollman, 1999.
\newblock {A modified version of the Cornell et al. force field with improved
  sugar pucker phases and helical repeat}.
\newblock \emph{J. Biomol. Struct. Dyn.} 16:845--862.

\bibitem[P{\'e}rez et~al.(2007)P{\'e}rez, March{\'a}n, Svozil, Sponer,
  Cheatham, Laughton, and Orozco]{Perez:2007ek}
P{\'e}rez, A., I.~March{\'a}n, D.~Svozil, J.~Sponer, T.~E. Cheatham, 3rd, C.~A.
  Laughton, and M.~Orozco, 2007.
\newblock {Refinement of the \texttt{AMBER} force field for nucleic acids:
  improving the description of $\alpha$/$\gamma$ conformers}.
\newblock \emph{Biophys. J.} 92:3817--3829.

\bibitem[Fonseca~Guerra et~al.(1999)Fonseca~Guerra, Bickelhaupt, Snijders, and
  Baerends]{FonsecaGuerra:1999ki}
Fonseca~Guerra, C., F.~M. Bickelhaupt, J.~G. Snijders, and E.~J. Baerends,
  1999.
\newblock {The nature of the hydrogen bond in DNA base pairs: the role of
  charge transfer and resonance assistance}.
\newblock \emph{Chemistry.} 5:3581--3594.

\bibitem[SantaLucia(1998)]{SantaLucia:1998uz}
SantaLucia, J., 1998.
\newblock {A unified view of polymer, dumbbell, and oligonucleotide DNA
  nearest-neighbor thermodynamics}.
\newblock \emph{Proc. Natl. Acad. Sci. USA.} 95:1460--1465.

\bibitem[Kumar et~al.(1992)Kumar, Rosenberg, Bouzida, Swendsen, and
  Kollman]{Kumar:1992bv}
Kumar, S., J.~M. Rosenberg, D.~Bouzida, R.~H. Swendsen, and P.~A. Kollman,
  1992.
\newblock {The weighted histogram analysis method for free-energy calculations
  on biomolecules. I. The method}.
\newblock \emph{J. Comput. Chem.} 13:1011--1021.

\bibitem[Hub et~al.(2010)Hub, de~Groot, and van~der Spoel]{Hub:2010ex}
Hub, J.~S., B.~L. de~Groot, and D.~van~der Spoel, 2010.
\newblock {\texttt{g\_wham}---a free weighted histogram analysis implementation
  including robust error and autocorrelation estimates}.
\newblock \emph{J. Chem. Theory Comput.} 6:3713--3720.

\bibitem[Olson et~al.(2001)Olson, Bansal, Burley, Dickerson, Gerstein, Harvey,
  Heinemann, Lu, Neidle, Shakked, Sklenar, Suzuki, Tung, Westhof, Wolberger,
  and Berman]{Olson:2001cf}
Olson, W.~K., M.~Bansal, S.~K. Burley, R.~E. Dickerson, M.~Gerstein, S.~C.
  Harvey, U.~Heinemann, X.-J. Lu, S.~Neidle, Z.~Shakked, H.~Sklenar, M.~Suzuki,
  C.-S. Tung, E.~Westhof, C.~Wolberger, and H.~M. Berman, 2001.
\newblock {A standard reference frame for the description of nucleic acid
  base-pair geometry}.
\newblock \emph{J. Mol. Biol.} 313:229--237.

\bibitem[Bustamante et~al.(1994)Bustamante, Marko, Siggia, and
  Smith]{Bustamante:1994wp}
Bustamante, C., J.~F. Marko, E.~D. Siggia, and S.~Smith, 1994.
\newblock {Entropic elasticity of $\lambda$-phage DNA}.
\newblock \emph{Science} 265:1599--1600.

\bibitem[{Harrison} et~al.(2015{\natexlab{a}}){Harrison}, {Romano},
  {Ouldridge}, {Louis}, and {Doye}]{2015arXiv150609005H}
{Harrison}, R.~M., F.~{Romano}, T.~E. {Ouldridge}, A.~A. {Louis}, and J.~P.~K.
  {Doye}, 2015.
\newblock {Coarse-grained modelling of strong DNA bending I: Thermodynamics and
  comparison to an experimental ``molecular vice''}.
\newblock \emph{arXiv:1506.09005} .

\bibitem[{Harrison} et~al.(2015{\natexlab{b}}){Harrison}, {Romano},
  {Ouldridge}, {Louis}, and {Doye}]{2015arXiv150609008H}
{Harrison}, R.~M., F.~{Romano}, T.~E. {Ouldridge}, A.~A. {Louis}, and J.~P.~K.
  {Doye}, 2015.
\newblock {Coarse-grained modelling of strong DNA bending II: Cyclization}.
\newblock \emph{arXiv:1506.09008} .

\bibitem[Protozanova et~al.(2004)Protozanova, Yakovchuk, and
  Frank-Kamenetskii]{Protozanova:2004bo}
Protozanova, E., P.~Yakovchuk, and M.~D. Frank-Kamenetskii, 2004.
\newblock {Stacked{\textendash}unstacked equilibrium at the nick site of DNA}.
\newblock \emph{J. Mol. Biol.} 342:775--785.

\bibitem[Yakovchuk et~al.(2006)Yakovchuk, Protozanova, and
  Frank-Kamenetskii]{Yakovchuk:2006bm}
Yakovchuk, P., E.~Protozanova, and M.~D. Frank-Kamenetskii, 2006.
\newblock {Base-stacking and base-pairing contributions into thermal stability
  of the DNA double helix.}
\newblock \emph{Nucleic Acids Res.} 34:564--574.

\bibitem[Forties et~al.(2009)Forties, Bundschuh, and Poirier]{Forties:2009kva}
Forties, R.~A., R.~Bundschuh, and M.~G. Poirier, 2009.
\newblock {The flexibility of locally melted DNA}.
\newblock \emph{Nucleic Acids Res.} 37:4580--4586.

\bibitem[Shimada and Yamakawa(1984)]{Shimada:1984gw}
Shimada, J., and H.~Yamakawa, 1984.
\newblock {Ring-closure probabilities for twisted wormlike chains. Application
  to DNA}.
\newblock \emph{Macromolecules.} 17:689--698.

\bibitem[Qu et~al.(2010)Qu, Tseng, Wang, Levine, and Zocchi]{Qu:2010gz}
Qu, H., C.-Y. Tseng, Y.~Wang, A.~J. Levine, and G.~Zocchi, 2010.
\newblock {The elastic energy of sharply bent nicked DNA}.
\newblock \emph{Europhys. lett.} 90:18003.

\bibitem[Qu et~al.(2011)Qu, Wang, Tseng, and Zocchi]{Qu:2011jz}
Qu, H., Y.~Wang, C.-Y. Tseng, and G.~Zocchi, 2011.
\newblock {Critical torque for kink formation in double-stranded DNA}.
\newblock \emph{Phys. Rev. X.} 1:021008.

\bibitem[Shroff et~al.(2005)Shroff, Reinhard, Siu, Agarwal, Spakowitz, and
  Liphardt]{Shroff:2005cw}
Shroff, H., B.~M. Reinhard, M.~Siu, H.~Agarwal, A.~Spakowitz, and J.~Liphardt,
  2005.
\newblock {Biocompatible force sensor with optical readout and dimensions of 6
  nm$^3$}.
\newblock \emph{Nano Lett.} 5:1509--1514.

\bibitem[Du et~al.(2007)Du, Kotlyar, and Vologodskii]{Du:2007be}
Du, Q., A.~Kotlyar, and A.~Vologodskii, 2007.
\newblock {Kinking the double helix by bending deformation}.
\newblock \emph{Nucleic Acids Res.} 36:1120--1128.

\bibitem[Vologodskii and D~Frank-Kamenetskii(2013)]{Vologodskii:2013il}
Vologodskii, A., and M.~D~Frank-Kamenetskii, 2013.
\newblock {Strong bending of the DNA double helix}.
\newblock \emph{Nucleic Acids Res.} 41:6785--6792.

\bibitem[Wiggins et~al.(2006)Wiggins, van~der Heijden, Moreno-Herrero,
  Spakowitz, Phillips, Widom, Dekker, and Nelson]{Wiggins:2006fu}
Wiggins, P.~A., T.~van~der Heijden, F.~Moreno-Herrero, A.~Spakowitz,
  R.~Phillips, J.~Widom, C.~Dekker, and P.~C. Nelson, 2006.
\newblock {High flexibility of DNA on short length scales probed by atomic
  force microscopy}.
\newblock \emph{Nat. Nanotechnol.} 1:137--141.

\bibitem[Mazur and Maaloum(2014)]{PhysRevLett.112.068104}
Mazur, A.~K., and M.~Maaloum, 2014.
\newblock DNA flexibility on short length scales probed by atomic force
  microscopy.
\newblock \emph{Phys. Rev. Lett.} 112:068104.

\bibitem[Demurtas et~al.(2009)Demurtas, Amzallag, Rawdon, Maddocks, Dubochet,
  and Stasiak]{Demurtas:2009ju}
Demurtas, D., A.~Amzallag, E.~J. Rawdon, J.~H. Maddocks, J.~Dubochet, and
  A.~Stasiak, 2009.
\newblock {Bending modes of DNA directly addressed by cryo-electron microscopy
  of DNA minicircles}.
\newblock \emph{Nucleic Acids Res.} 37:2882--2893.

\bibitem[Lanka{\v s} et~al.(2006)Lanka{\v s}, Lavery, and
  Maddocks]{Lankas:2006jz}
Lanka{\v s}, F., R.~Lavery, and J.~H. Maddocks, 2006.
\newblock {Kinking occurs during molecular dynamics simulations of small DNA
  minicircles.}
\newblock \emph{Structure} 14:1527--1534.

\bibitem[P{\'e}rez et~al.(2008)P{\'e}rez, Lanka{\v s}, Luque, and
  Orozco]{Perez:2008ft}
P{\'e}rez, A., F.~Lanka{\v s}, F.~J. Luque, and M.~Orozco, 2008.
\newblock {Towards a molecular dynamics consensus view of \textit{B}-DNA
  flexibility.}
\newblock \emph{Nucleic Acids Res.} 36:2379--2394.

\bibitem[Curuksu et~al.(2009)Curuksu, Zacharias, Lavery, and
  Zakrzewska]{Curuksu:2009cv}
Curuksu, J., M.~Zacharias, R.~Lavery, and K.~Zakrzewska, 2009.
\newblock {Local and global effects of strong DNA bending induced during
  molecular dynamics simulations.}
\newblock \emph{Nucleic Acids Res.} 37:3766--3773.

\bibitem[Harris et~al.(2008)Harris, Laughton, and Liverpool]{Harris:2008hc}
Harris, S.~A., C.~A. Laughton, and T.~B. Liverpool, 2008.
\newblock {Mapping the phase diagram of the writhe of DNA nanocircles using
  atomistic molecular dynamics simulations.}
\newblock \emph{Nucleic Acids Res.} 36:21--29.

\bibitem[Liverpool et~al.(2008)Liverpool, Harris, and
  Laughton]{Liverpool:2008db}
Liverpool, T.~B., S.~A. Harris, and C.~A. Laughton, 2008.
\newblock {Supercoiling and denaturation of DNA loops}.
\newblock \emph{Phys. Rev. Lett.} 100:238103.

\bibitem[Mitchell and Harris(2013)]{Mitchell:2013cp}
Mitchell, J.~S., and S.~A. Harris, 2013.
\newblock {Thermodynamics of writhe in DNA minicircles from molecular dynamics
  simulations}.
\newblock \emph{Phys. Rev. Lett.} 110:148105.

\bibitem[Yamakawa(1972)]{Yamakawa:1972jk}
Yamakawa, H., 1972.
\newblock {Statistical mechanics of wormlike chains. II. Excluded volume
  effects}.
\newblock \emph{J. Chem. Phys.} 57:2843--2854.

\bibitem[Lu and Olson(2008)]{Lu:2008kb}
Lu, X.-J., and W.~K. Olson, 2008.
\newblock {\texttt{3DNA}: a versatile, integrated software system for the
  analysis, rebuilding and visualization of three-dimensional nucleic-acid
  structures}.
\newblock \emph{Nat. Protoc.} 3:1213--1227.

\bibitem[Bussi et~al.(2007)Bussi, Donadio, and
  Parrinello]{:/content/aip/journal/jcp/126/1/10.1063/1.2408420}
Bussi, G., D.~Donadio, and M.~Parrinello, 2007.
\newblock {Canonical sampling through velocity rescaling}.
\newblock \emph{J. Chem. Phys.} 126:014101.

\bibitem[Parrinello(1981)]{Parrinello:1981it}
Parrinello, M., 1981.
\newblock {Polymorphic transitions in single crystals: a new molecular dynamics
  method}.
\newblock \emph{J. Appl. Phys.} 52:7182--7190.

\bibitem[Clowney et~al.(1996)Clowney, Jain, Srinivasan, Westbrook, Olson, and
  Berman]{Clowney:1996kr}
Clowney, L., S.~C. Jain, A.~R. Srinivasan, J.~Westbrook, W.~K. Olson, and H.~M.
  Berman, 1996.
\newblock {Geometric parameters in nucleic acids: nitrogenous bases}.
\newblock \emph{J. Am. Chem. Soc.} 118:509--518.

\bibitem[Horn(1987)]{Horn:1987hf}
Horn, B. K.~P., 1987.
\newblock {Closed-form solution of absolute orientation using unit
  quaternions}.
\newblock \emph{J. Opt. Soc. Am. A.} 4:629--642.

\bibitem[Torrie and Valleau(1977)]{Torrie:1977hs}
Torrie, G.~M., and J.~P. Valleau, 1977.
\newblock {Nonphysical sampling distributions in Monte Carlo free-energy
  estimation: umbrella sampling}.
\newblock \emph{J. Comput. Phys.} 23:187--199.

\bibitem[Cocco et~al.(2004)Cocco, Yan, L{\'e}ger, Chatenay, and
  Marko]{Cocco:2004fo}
Cocco, S., J.~Yan, J.-F. L{\'e}ger, D.~Chatenay, and J.~Marko, 2004.
\newblock {Overstretching and force-driven strand separation of double-helix
  DNA}.
\newblock \emph{Phys. Rev. E Stat. Nonlin. Soft Matter Phys.} 70:011910.

\end{thebibliography}


\clearpage

\setcounter{figure}{0}

\begin{figure}
\centering
\includegraphics{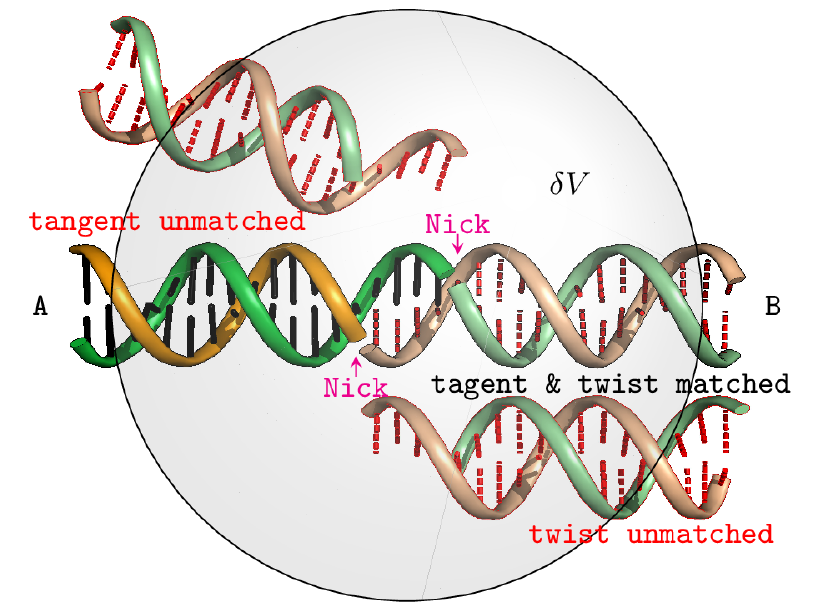}
\caption{$\Omega$-boundary condition in $j$-factor measurements. In
ligase-based DNA looping experiments, within the infinitesimal volume, $\delta
V$, around reference ``\texttt{A}'' end (\textit{with solid basepairing}), only
a subset of entered complimentary ``\texttt{B}'' ends (\textit{with dashed
basepairing}) can assemble into transiently stabilized hybridized \texttt{A-B}
ends, and chemically trapped by a subsequent ligation reaction. Under the
$\Omega$-boundary condition, it entails a $\left( 4 \pi \times 2
\pi\right)^{-1}$ factor. Tangent unmatched (\textit{top}) and twist unmatched
(\textit{bottom}) fragments, \texttt{B} ends are shown for comparison. Note
that two preexisting nicks (\textit{arrows}) are formed immediately after
hybridization, which may cause a violation of $\Omega$-boundary condition when
DNA is sharply bent.}
\label{fig:1Orientation}
\end{figure}

\begin{figure}
\centering
\includegraphics{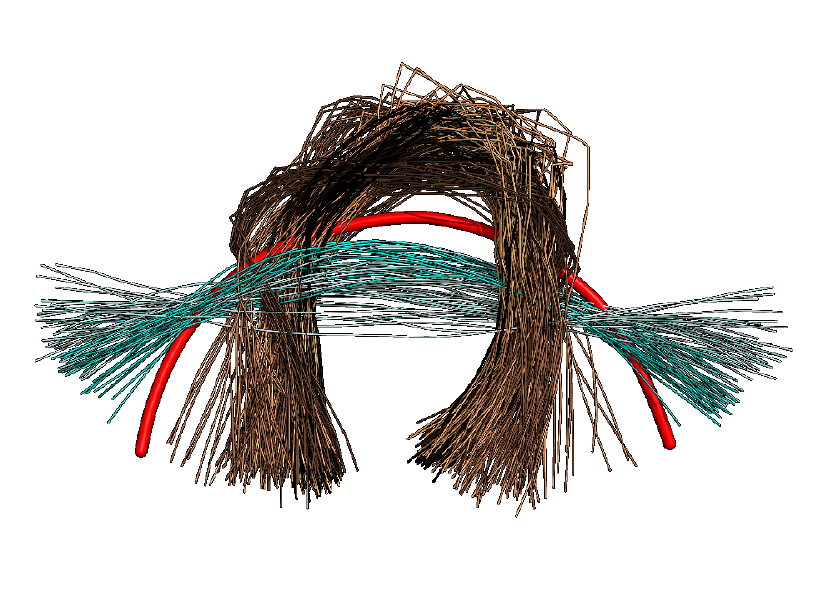}
\caption{Overview of distinctive DNA bending behaviors under weak and strong
spring constraints $\left\{\kappa; 0\right\}$. Above figure shows
superimpositions of DNA helical axes collected per ns in last $20$ ns for each
simulation. The fourteen independent MD simulations were all initiated from
same initial (represented by \textit{thick red helical axis}; atomic structure
is in Fig.~S1), and their corresponding stabilized centerlines are represented
(\textit{light cyan}) for weak spring constants $\kappa = 8.3$, $16.6$ pN/nm,
and (\textit{dark copper}) for strong bending $\kappa = 26.6$, $28.2$ (five
times), $29.0$, $31.5$, $33.2$, $41.5$, $49.8$, and $83.0$ pN/nm. When $\kappa
< 20.0$ pN/nm, the centerlines are uniformly bent and more straight than the
initial conformation. However, when $\kappa > 25.0$ pN/nm, the centerlines are
nonuniformly bent and more curved. Note that least curved backbones from
unconstrained simulations with $\kappa=0$ pN/nm are also included for
comparison.}
\label{fig:2Overview}
\end{figure}

\begin{figure}
\centering
\includegraphics{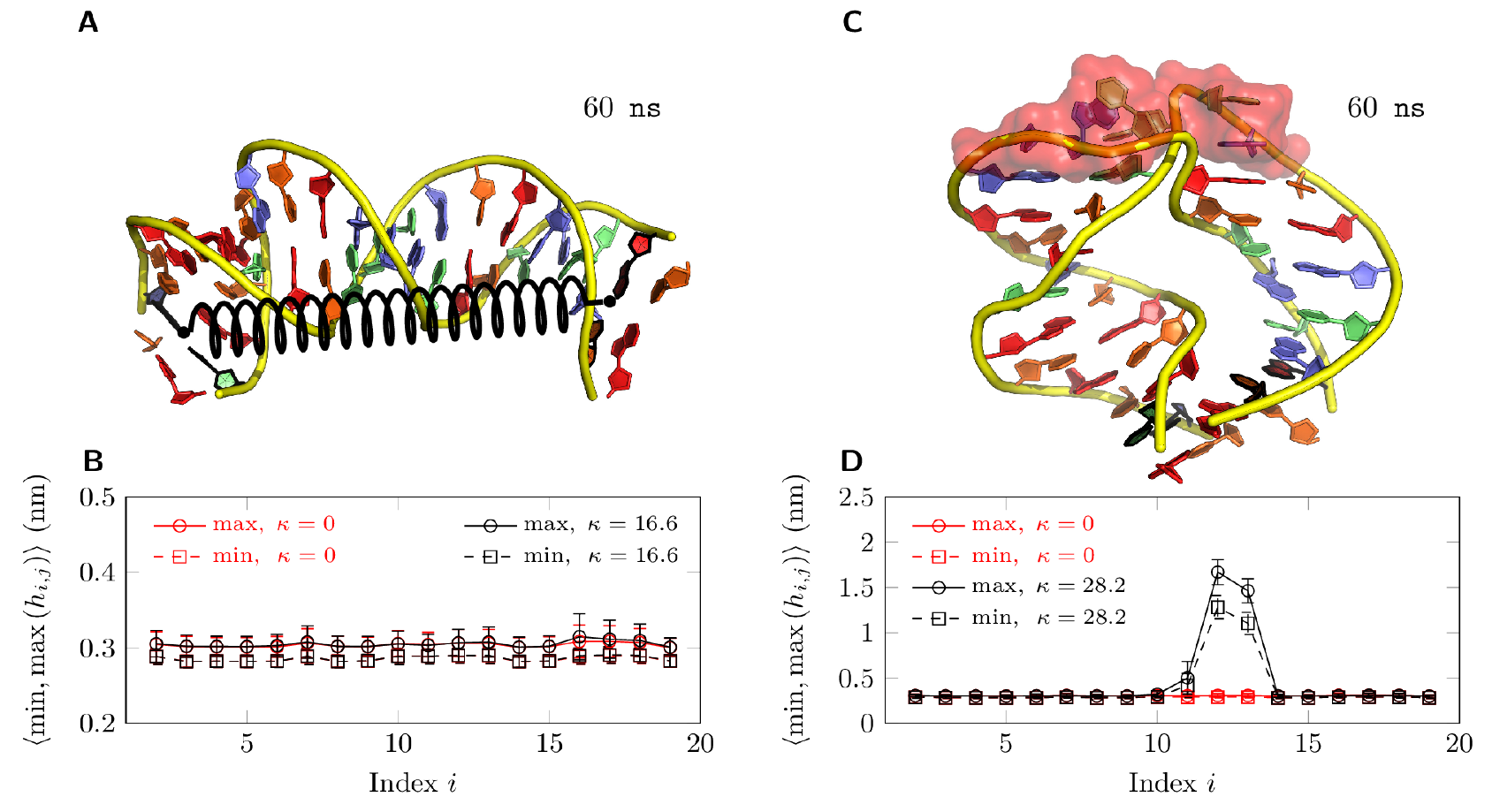}
\caption{Different DNA bending responses under weak and strong spring
constraints $\left\{ \kappa; 0 \right\}$. (\textbf{A}) A snapshot of a smoothly
bent DNA conformation at $t=60$ ns under a weak spring constant $\kappa=16.6$
pN/nm. (\textbf{B}) Corresponding hydrogen-bonding profile, $\left< \min, \max
\left( h_{i,j} \right) \right>$ plotted against $i$ values ($i=2,3,\cdots,19$)
averaged from the last $20$ of $70$ ns simulation. (\textbf{C}) A snapshot of a
severely bent DNA conformation at $60$ ns under a strong spring constant
$\kappa=28.2$ pN/nm, which contains a local basepair disruption in the middle.
(\textbf{D}) $\left< \min, \max \left( h_{i,j} \right) \right>$ averaged over
the last $20$ ns reveals three disrupted basepairs at $i=11, 12, 13$, which are
highlighted with the red surfaces in (\textbf{C}).}
\label{fig:3Compair}
\end{figure}

\begin{figure}
\centering
\includegraphics{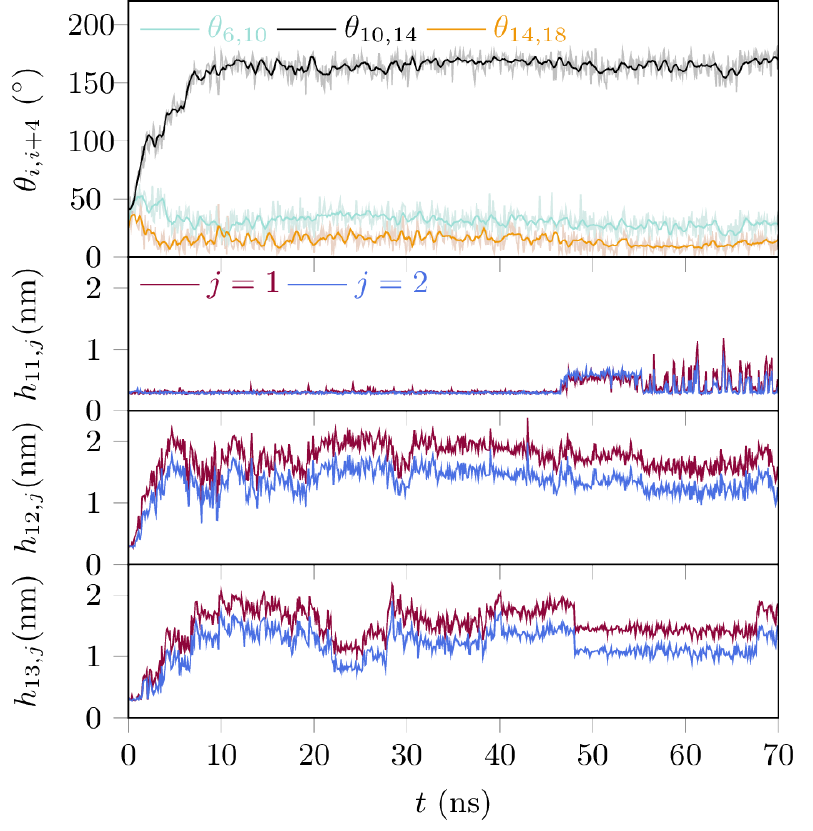}
\caption{The dynamics of local bending deformations and hydrogen-bond
disruptions under $\left\{ \kappa; 0 \right\}$ with $\kappa=28.2$ pN/nm over
$70$ ns. (\textit{Row $1$}) Evolution of $\theta_{10,14}$ enclosing three
basepairs at $i = 11, 12, 13$ disrupted during the simulation shows that kink
development around the region with disrupted DNA basepairs. The bending angle
evolution of two intact regions with same length, $\theta_{6,10}$ and
$\theta_{14,18}$, is shown for comparison. (\textit{Rows $2-4$}) Evolution of
$h_{i,j}$ for the three disrupted basepairs $i = 11, 12, 13$, which are all
\texttt{A$=$T} basepairs and involve two atom-atom distances each ($j=1$ and
$j=2$).}
\label{fig:4Dynamics}
\end{figure}

\begin{figure}
\centering
\includegraphics{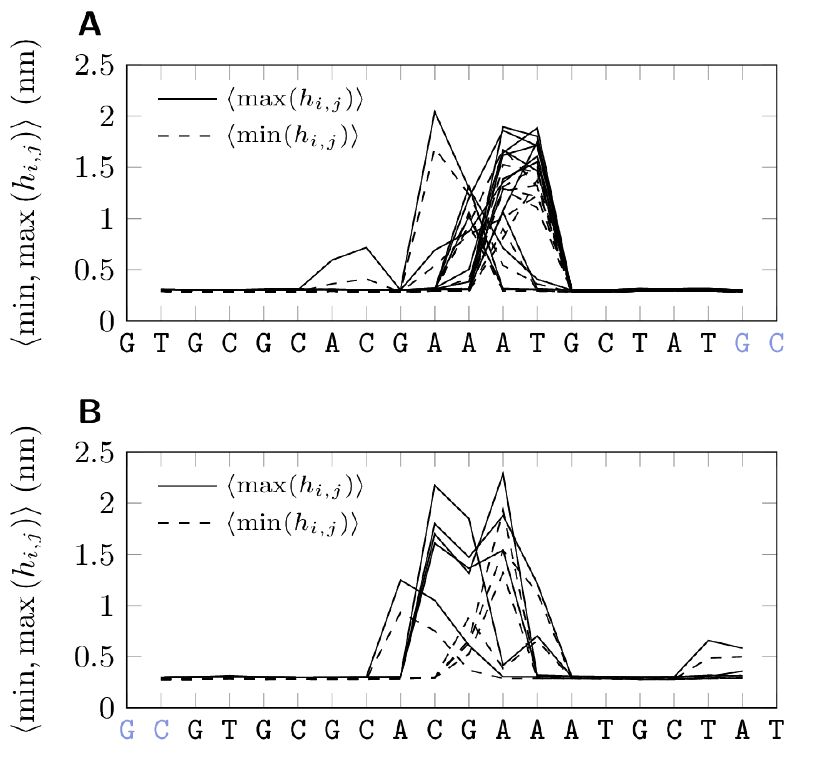}
\caption{Central localization of defects on different sequences.
Hydrogen-bonding profiles of DNA containing disrupted DNA basepairs: original
sequence $5'-\texttt{GTGCGCACGAAATGCTAT{\underline{GC}}}-3'$ and modified
sequence $5'-\texttt{{\underline{GC}}GTGCGCACGAAATGCTAT}-3'$. Overlay of
($\left< \min \left( h_{i,j} \right) \right>$, \textit{dashed lines}) and
($\left< \max \left( h_{i,j} \right) \right>$, \textit{solid lines}) along the
DNA sequence, averaged over the last $20$ ns for (\textbf{A}) twelve
independent simulations with the original sequence and (\textbf{B}) five
independent simulations with the modified sequence. All the hydrogen-bonding
profiles were obtained through constrained simulations ($\left\{ \kappa; 0
\right\}$), with various $\kappa > 25.0$ pN/nm (i.e., $\kappa = 26.6$, $28.2$
(five times), $29.0$, $31.5$, $33.2$, $41.5$, $49.8$, and $83.0$ pN/nm for the
original sequence; whereas $\kappa = 28.2$, $31.5$, $33.2$, $41.5$, and $49.8$
pN/nm for the modified sequence). The modified sequence was generated from the
original sequence by removing the tailing $5'-\texttt{\underline{GC}}-3'$ and
inserting it at the front, which offset the \texttt{AT}-rich region (i.e., its
$10^{\text{th}} - 13^{\text{th}}$ basepairs) away from its center.}
\label{fig:5DefectLocation}
\end{figure}

\begin{figure}
\centering
\includegraphics{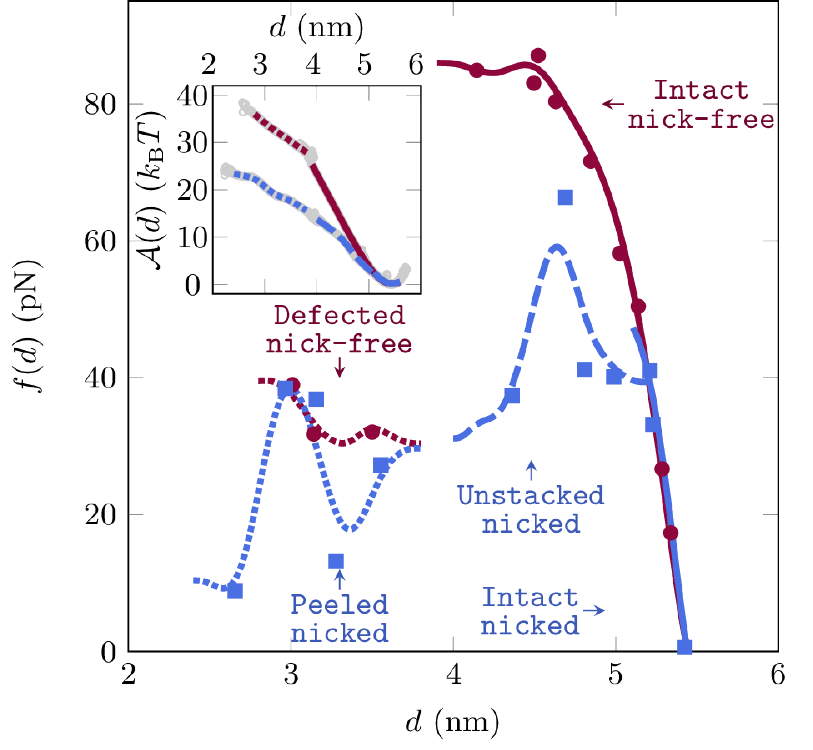}
\caption{The ${\cal A} \! \left( d\right)$ and $f \! \left( d\right)$ obtained
for various types of DNA at $300$ K. (\textit{Inset}) Smoothed ${\cal A} \!
\left( d\right)$, reference to global minimum state, for intact nick-free DNA
(\textit{dark-red solid line}), defect-containing nick-free DNA
(\textit{dark-red dotted line}), intact nicked DNA (\textit{light-blue solid
line}), unstacked nicked DNA (\textit{light-blue dashed line}), and peeled
nicked DNA (\textit{light-blue dotted line}). Main figure shows $f \!  \left(
d\right) = -\partial {\cal A}\!\left(d \right) / \partial d$ for different
types of DNA again were represented by different colors and line styles: intact
nick-free DNA (\textit{dark-red solid line}), defect-containing nick-free DNA
(\textit{dark-red dotted line}), intact nicked DNA (\textit{light-blue solid
line}), unstacked nicked DNA (\textit{light-blue dashed line}), and peeled
nicked DNA (\textit{light-blue dotted line}). For each type of DNA in the main
figure, the force values were directly read from the spring as well, which are
indicated by corresponding dots for nick-free DNA and corresponding squares for
nicked DNA. (\textit{Inset}, \textit{gray circles}) Discrete data obtained from
WHAM umbrella sampling analysis that were used to produce continuous ${\cal A}
\! \left( d\right)$ by cubic spline interpolation.}
\label{fig:6Profile}
\end{figure}

\begin{figure}
\centering
\includegraphics{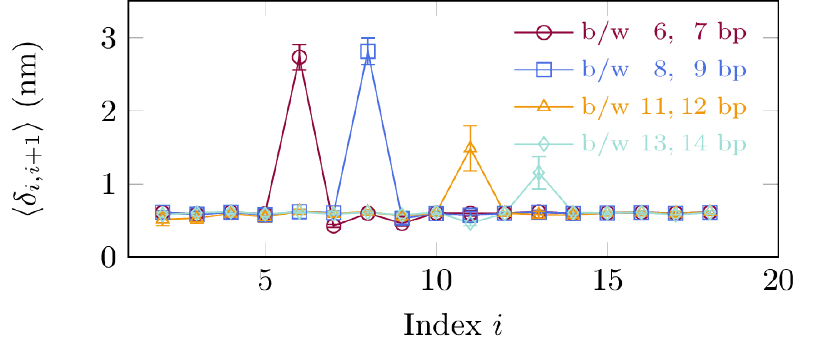}
\caption{Interbase distance profiles for the four nicked DNAs under a spring
constraint of $\left\{ 28.2; 0 \right\}$. The interbase distance profiles,
$\left< \delta_{i,i+1} \right>$ ($i=2,3,\cdots,18$) measure the averaged
distances between adjacent $\text{C4'}$ atoms of $i^{\text{th}}$ and ${\left( i
+ 1 \right)}^{\text{th}}$ basepairs on the entire top strand of DNAs in the
four independent simulations with nick right after the $6^{\text{th}}$,
$8^{\text{th}}$, $11^{\text{th}}$, and $13^{\text{th}}$ basepairs. The dramatic
increase in $\left< \delta_{i,i+1} \right>$ in the corresponding
nick-containing simulations reveals that disruptions of basepairs occurred at
nicked sites. Note that $\text{C4'}$ atoms of deoxyribose are part of the DNA
sugar-phosphate backbone.}
\label{fig:7NickDefect}
\end{figure}

\begin{figure}
\centering
\includegraphics{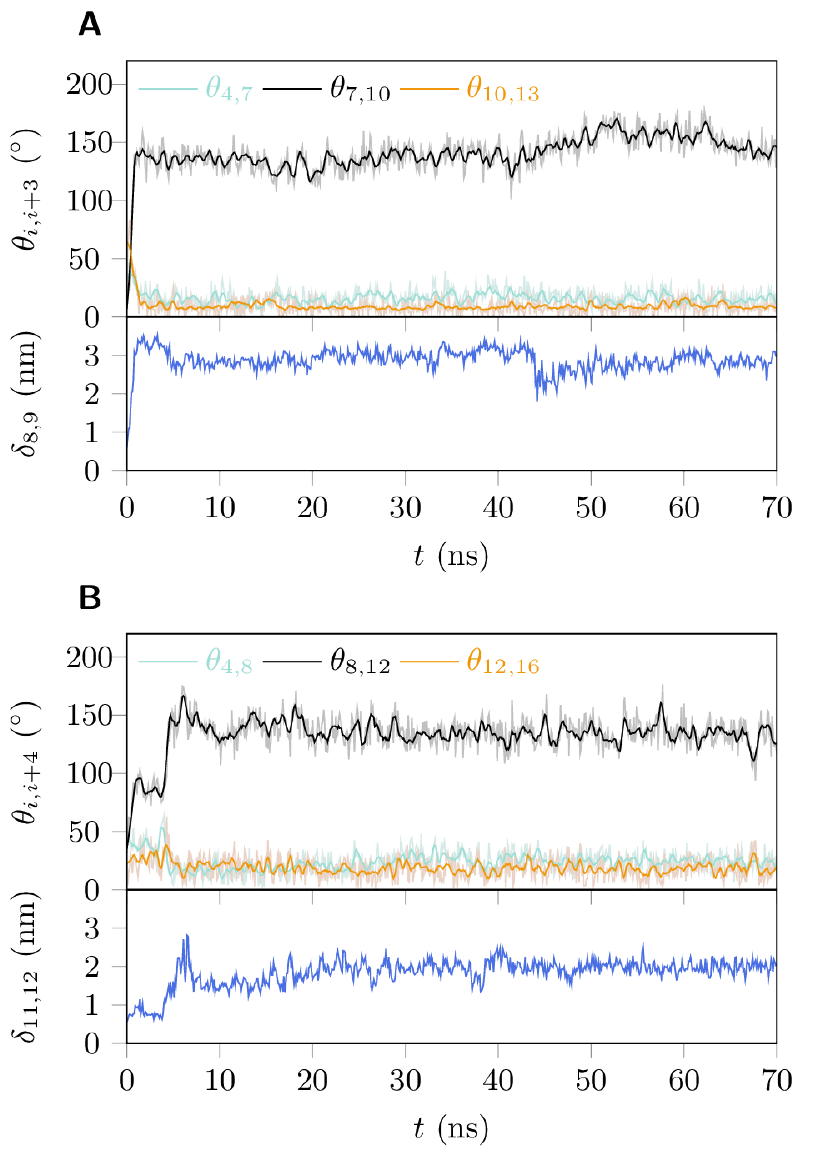}
\caption{The dynamics of local bending deformations and basepair separations at
nicked sites under a spring constraint of $\left\{28.2; 0 \right\}$ over $70$
ns. (\textbf{A}) (\textit{Row $1$}) Evolution of $\theta_{7,10}$, enclosing the
nicked site between the $8^{\text{th}}$ and $9^{\text{th}}$ basepairs, which
shows kink development around the unstacked region. The bending angle evolution
of two intact regions with same length, $\theta_{4,7}$ and $\theta_{10,13}$, is
shown for comparison. (\textit{Row $2$}) Evolution of $\delta_{8,9}$ indicates
basepair separation at nicked sites. (\textbf{B}) Similar dynamics of kink
development ($\theta_{8,12}$), bending relaxation ($\theta_{4,8}$ and
$\theta_{12,16}$), and basepair separation ($\delta_{11,12}$) for the peeled
DNA with nick between $11^{\text{th}}$ and $12^{\text{th}}$ basepair.}
\label{fig:8NickDyn}
\end{figure}

\begin{figure}
\centering
\includegraphics{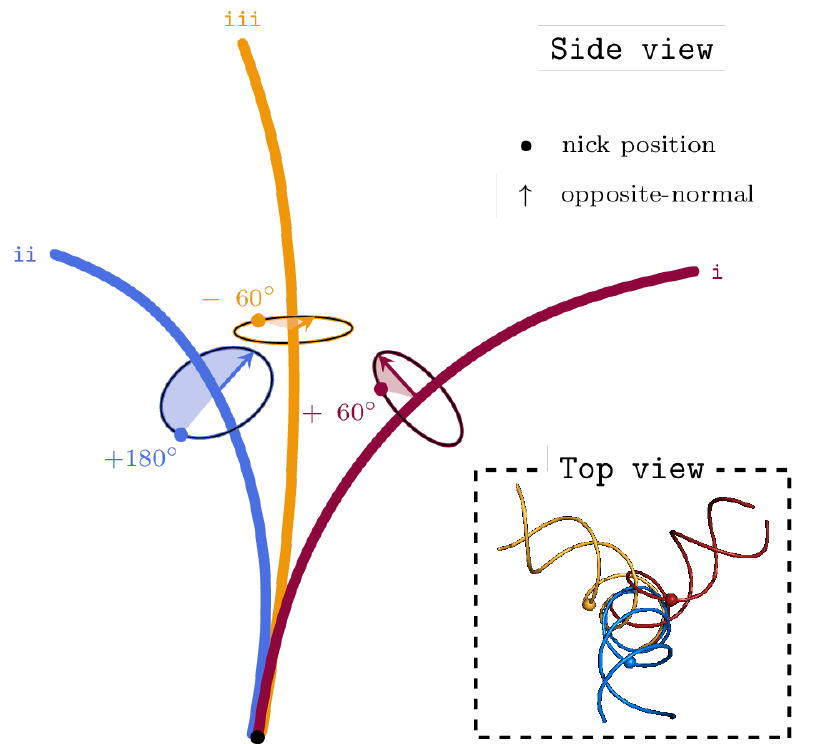}
\caption{Initial conformations for nicked and nick-free DNA bent into different
directions. The first basepairs are superimposed; therefore, the initial
conformations have the same starting orientation. The three DNA molecules are
bent uniformly outward in three distinctive directions, denoted \rmnum{1},
\rmnum{2}, and \rmnum{3}, with their end-to-end distances projected onto the
first basepair plane evenly separated. (\textit{Side view}) At the particular
location corresponding to where a nick is introduced, a local polar coordinate
is defined with the opposite-normal direction as its polar axis
(\textit{indicated with arrows}). The nick positions (\textit{indicated with
dots}) in the DNAs are $\pm 60^\circ$ and $+180^\circ$ in the corresponding
local polar coordinates. (\textit{Inset}, \textit{top view}) The three DNA
duplexes with spheres denoting the phosphate groups that are deleted in the
nicked DNA on Strand \Rmnum{1} between the $11^{\text{th}}$ and
$12^{\text{th}}$ basepairs. The initial bending is controlled by tilts and
rolls of the basepairs provided in Table~S1.}
\label{fig:9DBIni}
\end{figure}

\begin{figure}
\centering
\includegraphics{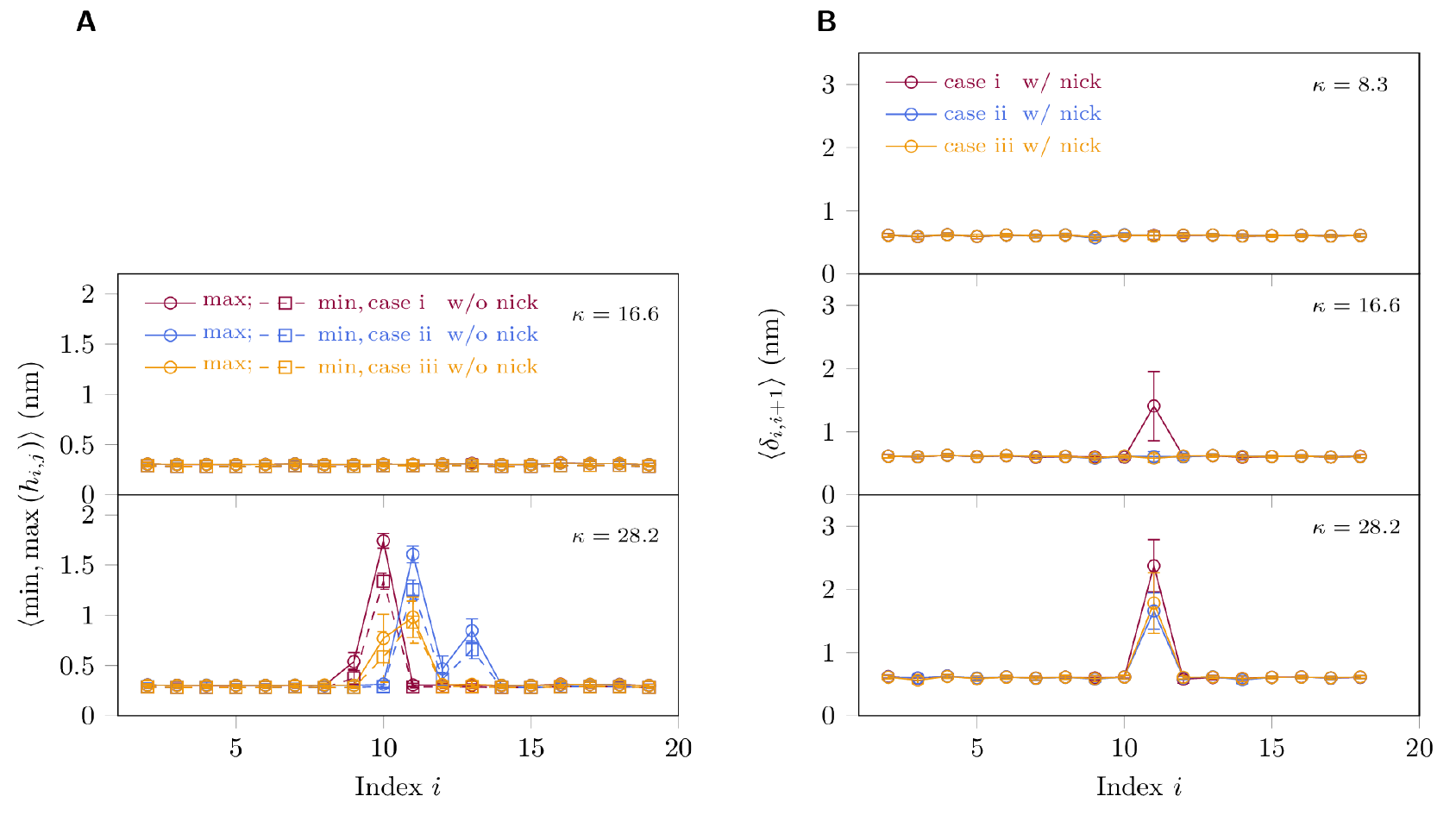}
\caption{Effects of direction of bending on defect excitation in three distinct
directions \rmnum{1}, \rmnum{2}, and \rmnum{3}. DNA molecules without and with
nicks were forcibly bent toward distinctive directions using various spring
constraints of $\left\{ \kappa; 0 \right\}$. (\textbf{A}) The hydrogen-bonding
profiles of nick-free DNA, ($\left< \min \left( h_{i,j} \right) \right>$,
\textit{dashed lines}) and ($\left< \max \left( h_{i,j} \right) \right>$,
\textit{solid lines}) along the DNA sequence averaged in $50-70$ ns
trajectories for different bending directions under constraints of
$\kappa=16.6$ (\textit{top}) and $28.2$ (\textit{bottom}) pN/nm. (\textbf{B})
Interbase distance profiles ($\left< \delta_{i,i+1} \right>$) between adjacent
$\text{C4'}$ atoms on Strand \Rmnum{1} for the nick-containing DNA, averaged in
$50-70$ ns trajectories for the three bending directions under three spring
constants of $\kappa=8.3$ (\textit{top}), $16.6$ (\textit{middle}), and $28.2$
(\textit{bottom}) pN/nm.}
\label{fig:10Isotropicity}
\end{figure}

\begin{figure}
\centering
\includegraphics{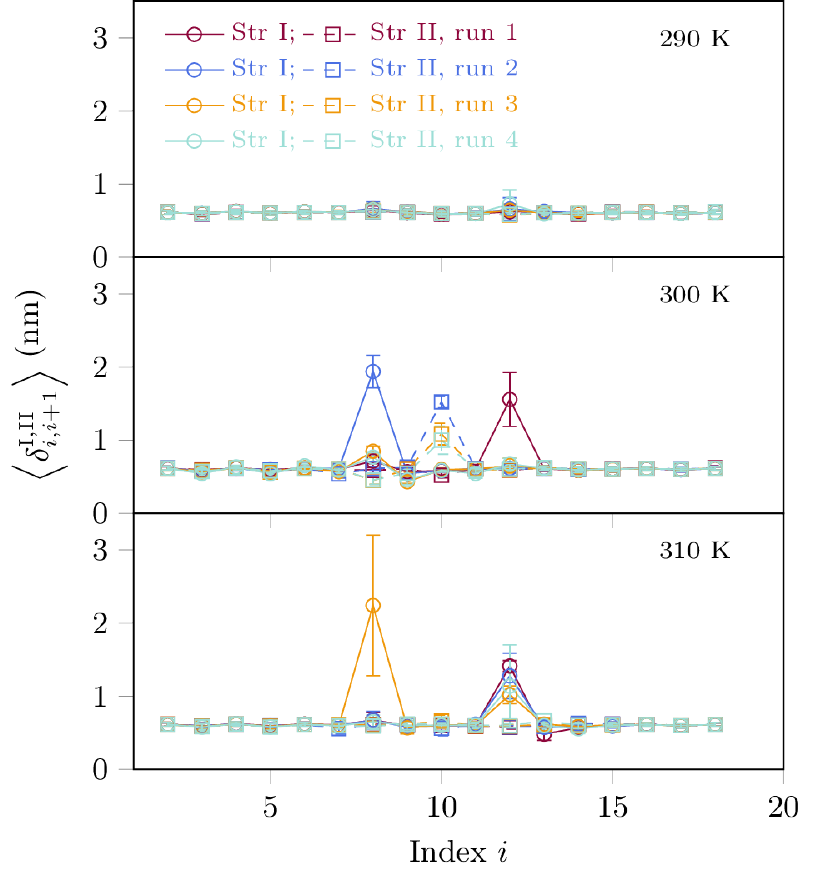}
\caption{Effects of temperature on nick-dependent defect excitation. DNA
molecules with triple nicks were constrained by the spring of $\left\{ 248.9;
4.20 \right\}$. Four independent $50$ ns simulations were performed for each
indicated temperature. The panels show the interbase distance profiles for both
strands along the DNA averaged in the last $20$ ns of each simulation: $\left<
\delta_{i,i+1}^\text{\Rmnum{1}} \right>$ (\textit{solid}) and $\left<
\delta_{i,i+1}^\text{\Rmnum{2}} \right>$ (\textit{dashed}), where $i$ denotes
the basepair index, and superscripts \Rmnum{1} and \Rmnum{2} denote the top and
bottom strands, respectively.}
\label{fig:11TempC4}
\end{figure}

\end{document}


\title{Supporting Material for: Revisiting the Anomalous Bending Elasticity of Sharply Bent DNA}

\author{Peiwen Cong}
\affiliation{Mechanobiology Institute, National University of Singapore, Singapore 117411, Singapore}
\affiliation{Computation and Systems Biology, Singapore-MIT Alliance, Singapore 117576, Singapore}
\affiliation{Department of Physics, National University of Singapore, Singapore 117551, Singapore}
\author{Liang Dai}
\affiliation{BioSystems and Micromechanics IRG, Singapore-MIT Alliance for Research and Technology Centre, Singapore 138602, Singapore}
\author{Hu Chen}
\affiliation{Department of Physics, Xiamen University, Xiamen, Fujian 361005, China}
\author{Johan R. C. van der Maarel}
\affiliation{Department of Physics, National University of Singapore, Singapore 117551, Singapore}
\author{Patrick S. Doyle}
\affiliation{Department of Chemical Engineering, Massachusetts Institute of Technology, Cambridge, Massachusetts 02139, USA}
\affiliation{BioSystems and Micromechanics, Singapore-MIT Alliance for Research and Technology Centre, Singapore 138602, Singapore}
\author{Jie Yan}
\email[Email address: ]{phyyj@nus.edu.sg}
\affiliation{Department of Physics, National University of Singapore, Singapore 117551, Singapore}
\affiliation{Mechanobiology Institute, National University of Singapore, Singapore 117411, Singapore}
\affiliation{Centre for BioImaging Sciences, National University of Singapore, Singapore 117557, Singapore}
\affiliation{BioSystems and Micromechanics, Singapore-MIT Alliance for Research and Technology Centre, Singapore 138602, Singapore}
\maketitle

\newpage
\section{Supporting Figures}

\begin{figure}[!htb]
\centering
\includegraphics{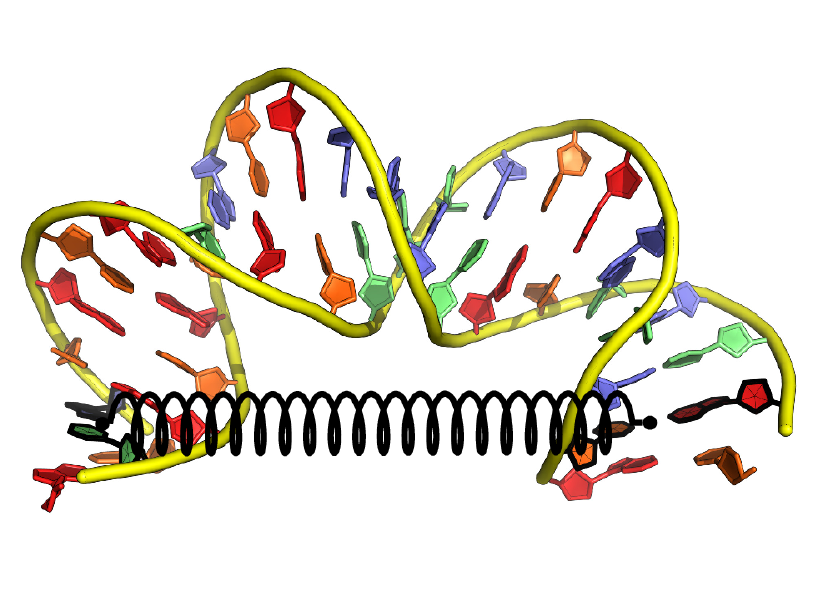}
\caption{\textbf{Initial smoothly bent DNA conformation generated by
\texttt{X3DNA}.} This initial conformation has an overall bending angle of
$\sim 160^{\circ}$. A constraining spring is connected to the bases of second
and second-last basepairs (\textit{black outlines}) to actively pull the DNA
ends inward. Note that the nucleotides are colored by sequence, \texttt{A} in
blue, \texttt{T} in green, \texttt{G} in red and \texttt{C} in orange, while
sugar-phosphate backbones are colored yellow.}
\label{fig:FIGS1}
\end{figure}

\begin{figure}[!htb]
\centering
\includegraphics{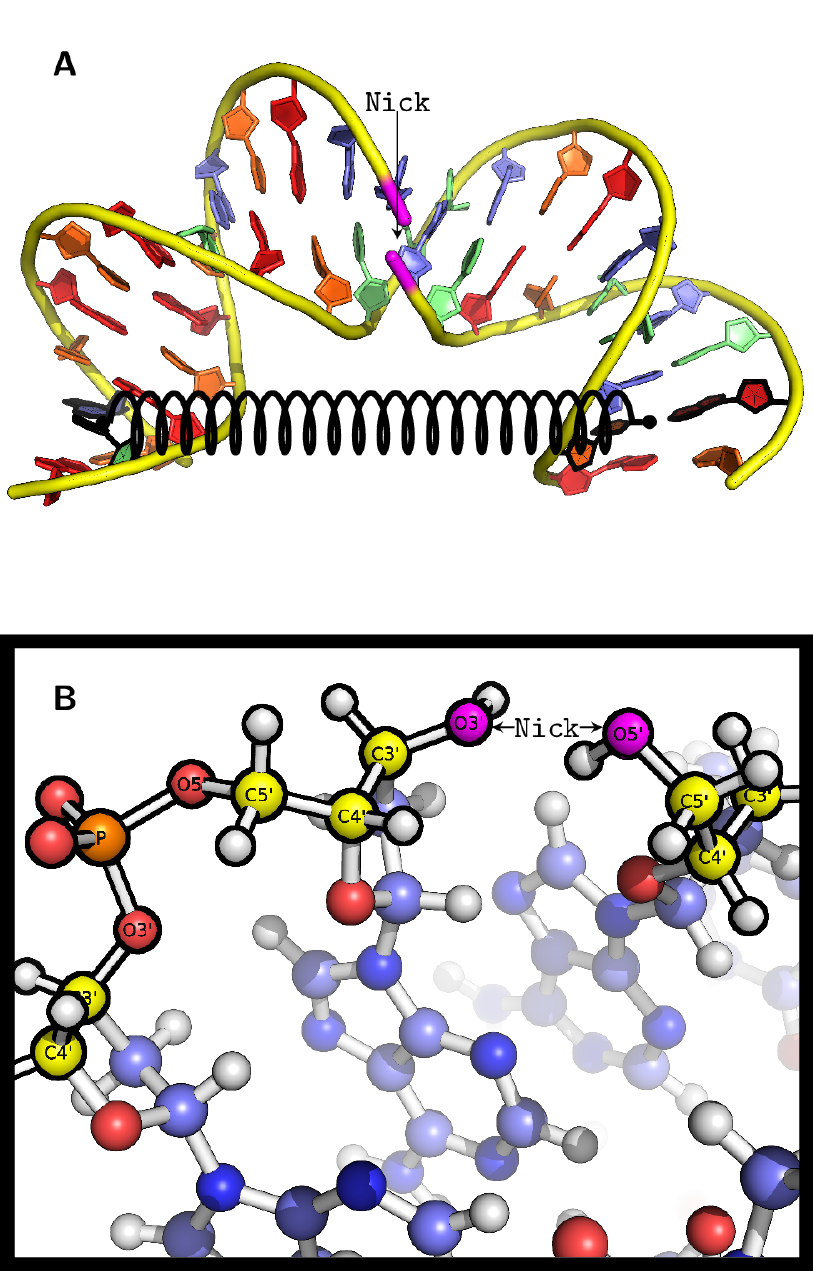}
\caption{\textbf{Nicked DNA construct with nick after the $11^\text{th}$
basepair.} (\textbf{A}) The initial smoothly bent DNA containing a nick between
the $11^{\text{th}}$ and $12^{\text{th}}$ basepairs in Strand \Rmnum{1}
(\textit{arrow}). (\textbf{B}) Magnification of the nicked site, where the
phosphodiester bonds were cleaved and the entire phosphate group was removed,
leaving the $\text{O3'}$ and $\text{O5'}$ atoms (\textit{magenta}) hydrolyzed.
The backbone carbon atoms are colored yellow, phosphate atoms are colored
orange, and oxygen atoms are colored red. The parameters describing resulting
terminal nucleotide residues ($\text{-OH}$) at the nick are included in Parm99
force field with ParmBSC0 corrections.}
\label{fig:NickConstruct}
\end{figure}

\begin{figure}[!htb]
\centering
\includegraphics{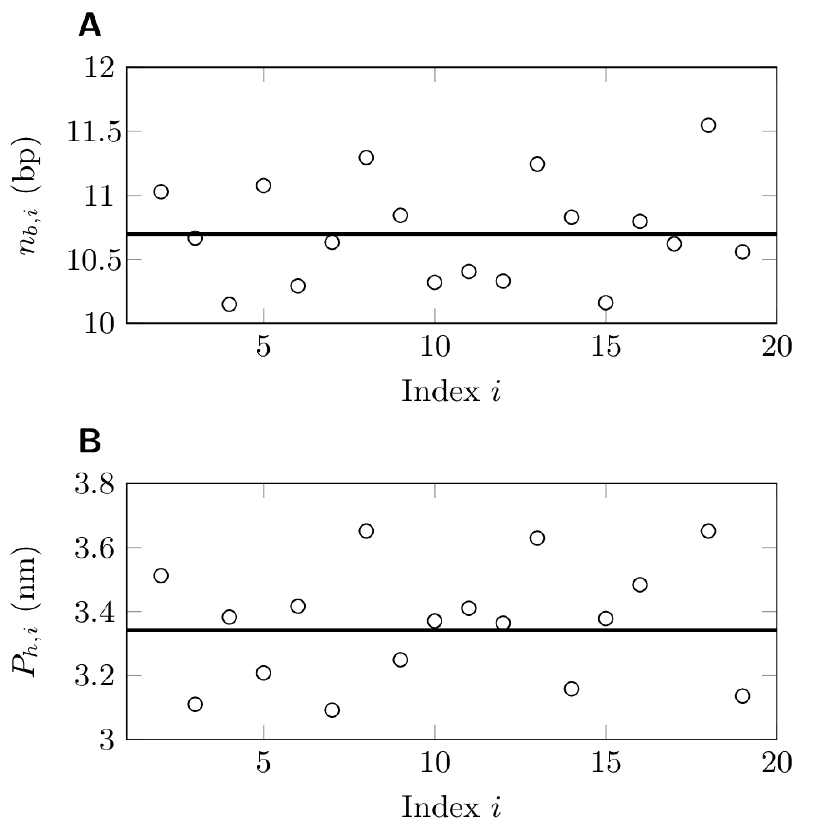}
\caption{\textbf{Helical parameters for \textit{B}-DNA without constraints.}
(\textbf{A}) Helical repeat, $n_{b,i}$ and (\textbf{B}) helical pitch,
$P_{h,i}$ along DNA (\textit{open circles}) were calculated using average twist
and rise at particular site $i$ over the last $20$ ns out of $70$ ns
unconstrained simulation. The global mean values are $10.70 \pm 1.53$ bp and
$3.34 \pm 0.67$ nm respectively (\textit{horizontal lines}), obtained through
$n_b={2\pi} / \left< \Omega \right>$ and $P_h={ 2\pi \left< D_z \right>} /
\left< \Omega \right>$, where $\Omega$ is twist angle, $D_z$ is rise per
basepair step. Note that the values after $\pm$ sign are the corresponding
standard deviations of uncorrelated structure representatives. $n_{b,i}$ and
$P_{h,i}$ are all around their global mean values, which indicates the
homogeneity of unconstrained DNA.}
\label{fig:Helical}
\end{figure}

\begin{figure}[!htb]
\centering
\includegraphics{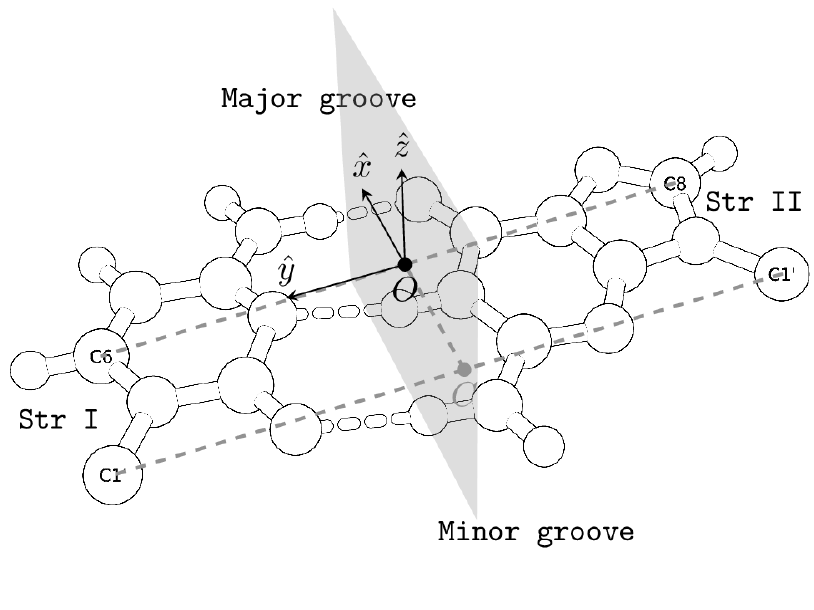}
\caption{\textbf{Basepair reference frame.} Reference frame for ideal
Watson-Crick basepair using \texttt{C$\equiv$G} basepair as an example, and
only complementary bases are shown. The coordinate is defined by four atoms,
\text{C1'}, \text{C6} from pyrimidine nucleotides (\texttt{C} and \texttt{T}),
and \text{C1'}, \text{C8} from purine nucleotides (\texttt{G} and \texttt{A}).
The plane (\textit{grey}), which is the perpendicular bisector of the line
segment $\overline{ \left( \text{C1'}~\text{C1'} \right) }$ at the midpoint
$C$, intersects with the line segment $\overline{ \left( \text{C6}~\text{C8}
\right) }$ at $O$. \textit{x}-axis directs from $C$ to $O$. \textit{y}-axis is
parallel to $\overline{ \left( \text{C1'}~\text{C1'} \right) }$, pointing
towards the Strand \Rmnum{1}. \textit{z}-axis is $\hat{z} = \hat{x} \times
\hat{y}$.}
\label{fig:RefFrame}
\end{figure}

\begin{figure}[!htb]
\centering
\includegraphics{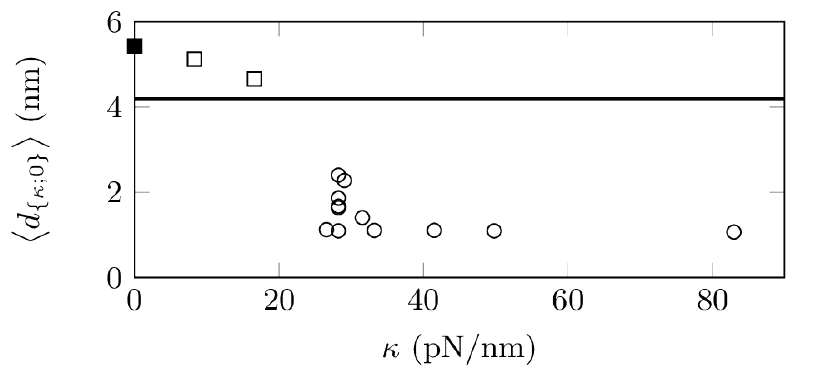}
\caption{\textbf{Mean values of end-to-end distances $\left< d_{\left\{\kappa;
0\right\}} \right>$ under various spring constrains.} They were averaged over
last $20$ ns for each simulation. $\left< d \right>$ with $\kappa=0$ pN/nm from
the unconstrained simulation ($\blacksquare$) is shown as control. $\left<
d_{\left\{\kappa; 0\right\}} \right>$ with $\kappa < 20.0$ pN/nm ($\square$)
are longer than $d_{\text{ini}}$ (\textit{horizontal line}), shorted than
control, and negatively correlated with $\kappa$. $\left< d_{\left\{\kappa;
0\right\}} \right>$ with $\kappa > 25.0$ pN/nm ($\ocircle$) are much shorter
than $d_{\text{ini}}$, about DNA diameter, and uncorrelated with $\kappa$.}
\label{fig:EtETotal}
\end{figure}

\begin{figure}[!htb]
\centering
\includegraphics{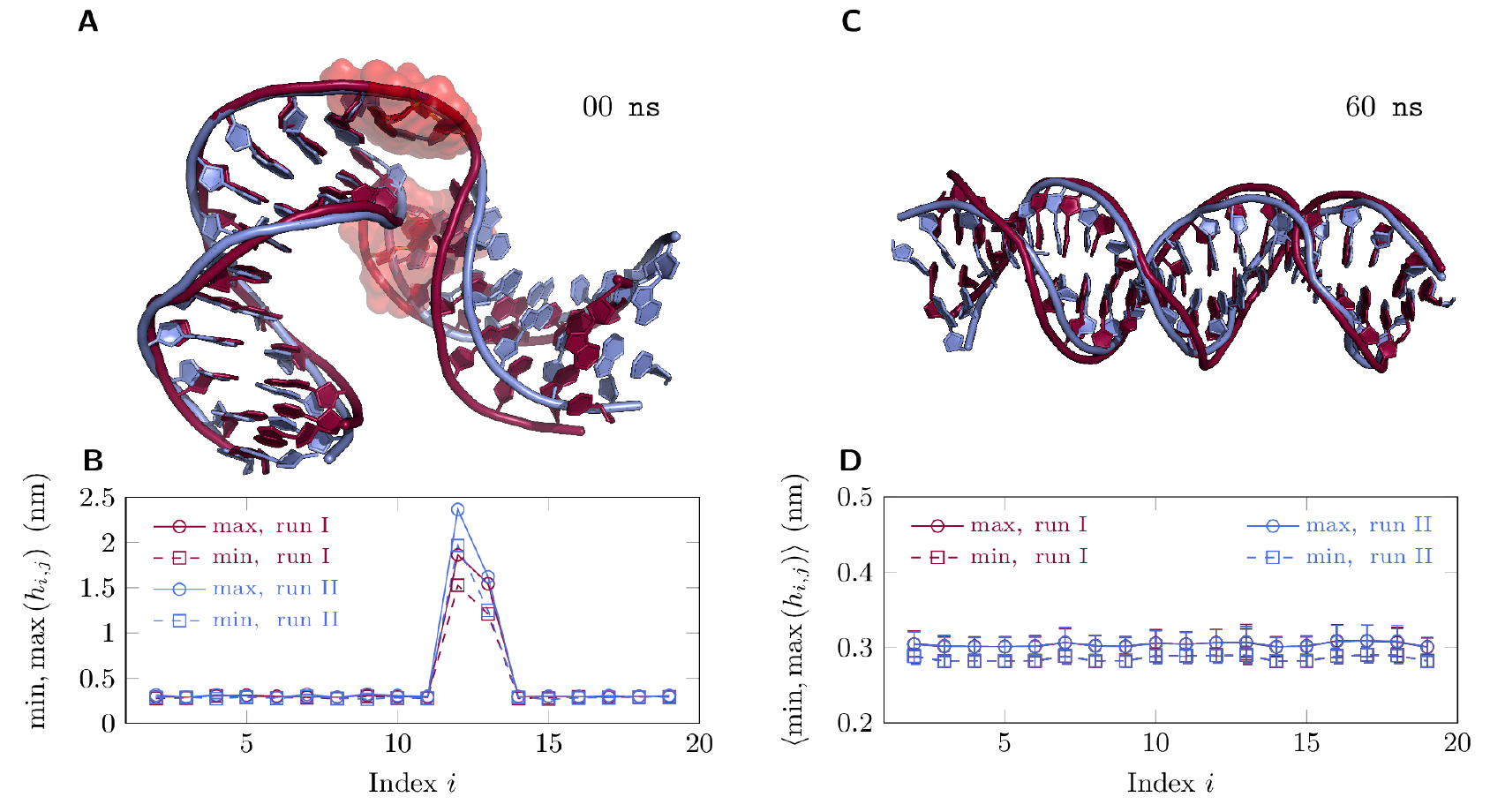}
\caption{\textbf{Reversibility of DNA defects.} Two $70$ ns simulations were
conducted to check the reversibility of defects once the sharp bending
constraint was removed. Runs \Rmnum{1} (\textit{dark red}) and \Rmnum{2}
(\textit{light blue}) were started from two different defect-containing DNAs
induced by sharp bending. (\textbf{A}) The initial atomic structures of runs
\Rmnum{1} and \Rmnum{2} both with defected $12^\text{th}$ and $13^\text{th}$
basepairs highlighted with the red surfaces. (\textbf{B}) Their corresponding
hydrogen-bonding distances, $\min, \max \left(h_{i,j} \right)$ plotted against
$i$ ($i=2, 3, \cdots, 19$) at $0$ ns. (\textbf{C}) The snapshots taken at $60$
ns after the simulations began, which show straightened \textit{B}-form
conformations. (\textbf{D}) The resulting hydrogen-bonding profiles, $\left<
\min, \max \left(h_{i,j} \right) \right>$ along DNA averaged in $50-70$ ns
trajectories overlap with that of control, which was obtained from previous
unconstrained simulation for intact \textit{B}-form DNA (\textit{black}).
Thus, the sharp bending induced defects are transient, and are able to restore
into \textit{B}-form given that the bending constraint is removed.}
\label{fig:Reversibility}
\end{figure}

\begin{figure}[!htb]
\centering
\includegraphics{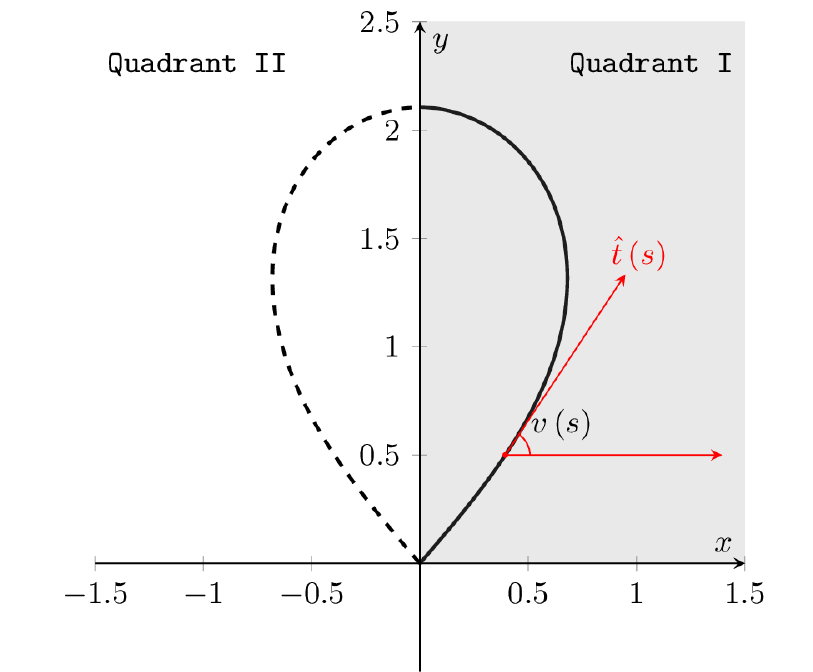}
\caption{\textbf{Energy minimal rigid path of short DNA fragment.} Without
considering thermal fluctuations, a planar looped rigid homogeneous polymer
under free boundary conditions forms a symmetric path, whose energy minimal
conformation assumes a teardrop shape. By defining the angle between Quadrant
\Rmnum{1} unit tangent vector $\hat{t} \! \left( s \right)$ and $x$-axis as
$\upsilon \! \left( s \right)$, it relates to curvature as, $L^2 \left(
{\partial{\hat{t} \! \left( s \right)} } / { \partial{s} } \right)^2 = -
\lambda \cos \left( \upsilon \! \left( s \right) \right) + c$, where $L$ is the
contour length, $\lambda > 0$ is Lagrange multiplier, and $c > 0$ is
integration constant \cite{Yamakawa:1972jk}. This implies maximized curvature
at its center, as $\upsilon \left( L / 2 \right) = \pi$ by symmetry.  Here, the
contour length was set to be $L=d_\text{e}=5.43$ nm same as our simulated
equilibrium length of $20$ bp DNA, whose two meeting termini make an angle of
$\theta = 81^\circ24^\prime$ in this teardrop shape.}
\label{fig:RigidPath}
\end{figure}

\begin{figure}[!htb]
\centering
\includegraphics{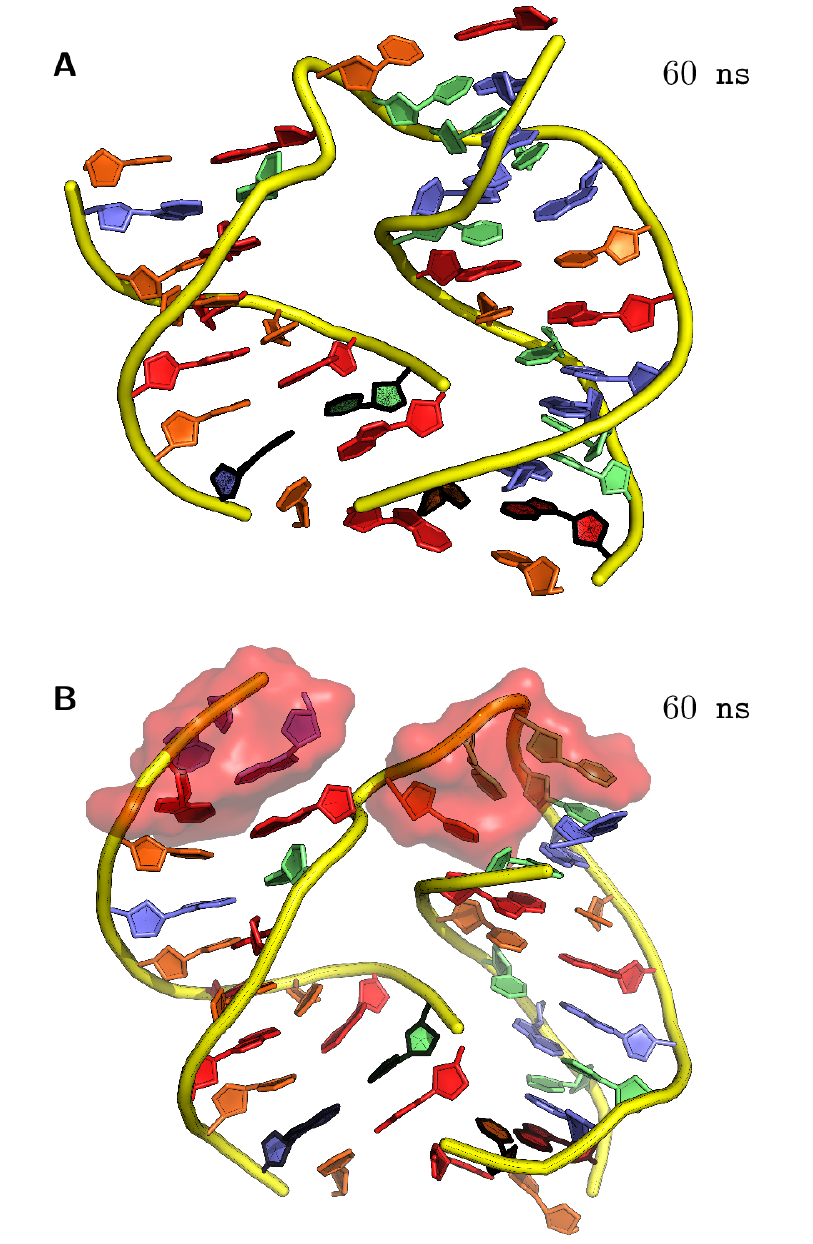}
\caption{\textbf{Atomic structures for unstacked and peeled nick-containing
DNAs.} (\textbf{A}) A snapshot at $t=60$ ns demonstrates an unstacked case,
which was extracted from the $\left\{28.2;0\right\}$-constrained simulation of
nicked DNA with nick on Strand \Rmnum{1} between the $8^\text{th}$ and
$9^\text{th}$ basepairs. (\textbf{B}) A snapshot at $t=60$ ns shows a peeled
case, which was clipped from the $\left\{28.2;0\right\}$-constrained trajectory
of nicked DNA with nick on Strand \Rmnum{1} between the $11^\text{th}$ and
$12^\text{th}$ basepairs. The red surfaces indicate the peeled Strand \Rmnum{1}
from the nick and their unpaired complementary bases at the disrupted
$9^\text{th}$, $10^\text{th}$, and $11^\text{th}$ basepairs.}
\label{fig:NickView}
\end{figure}

\begin{figure}[!htb]
\centering
\includegraphics{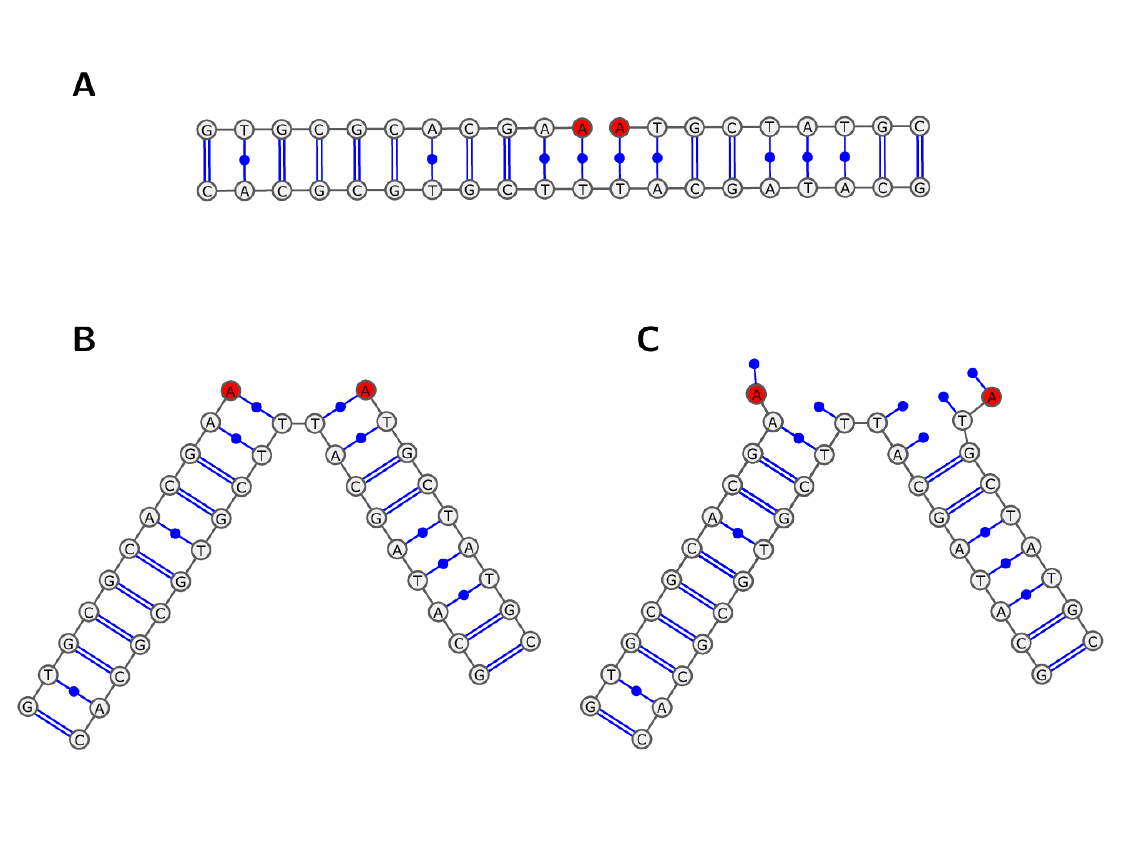}
\caption{\textbf{Illustrations for nicked DNA with different categories of
noncovalent topologies.} (\textbf{A}) Type A shows the intact nicked DNA with
both intact hydrogen-bonding and basepair-stacking. (\textbf{B}) Type B
represents the unstacked nicked DNA with disrupted basepair stacking only at
nicked position. (\textbf{C}) Type C indicates a particular case of the peeled
nicked DNA with both nicked ends split, resulting in both disrupted stacking
and basepairing around the nicked site. These illustrations use nonhelical
representations of nicked DNA with nicks after the $11^\text{th}$ basepair as
examples.}
\label{fig:NickCase}
\end{figure}

\begin{figure}[!htb]
\centering
\includegraphics{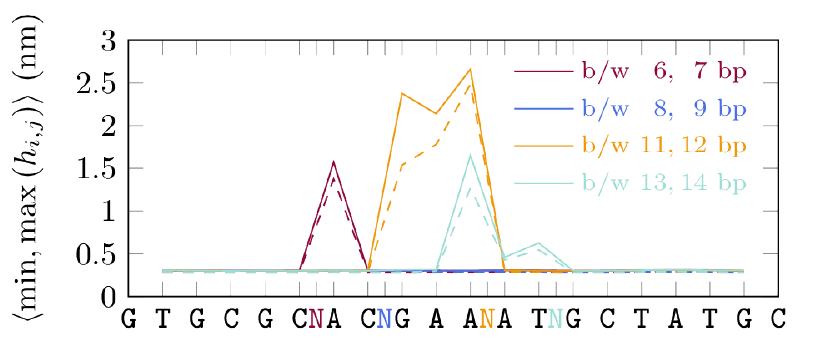}
\caption{\textbf{Hydrogen-bonding profiles for the four nicked DNAs under
$\left\{28.2;0\right\}$.} The hydrogen-bonding profiles, $\left< \min, \max
\left(h_{i,j} \right) \right>$ plotted against basepair index $i$ averaged over
the last $20$ ns of $70$ ns trajectories for four independent simulations with
nick right after the $6^{\text{th}}$, $8^{\text{th}}$, $11^{\text{th}}$, and
$13^{\text{th}}$ basepairs. Although their \text{C4'} interbase distance
profiles in main text have already indicated the presences of basepair-stacking
disruptions at the nicked sites, these hydrogen-bonding profiles further reveal
the existence of two distinctive types of disruptions: clean unstacking at
nicked site (i.e., with totally intact hydrogen-bonding) in the case of nick
after the $8^{\text{th}}$ basepair; and unstacking accompanied with peeling
from nicked sites (i.e., with locally disrupted hydrogen-bonding) in the case
of nicks after the $6^{\text{th}}$, $11^{\text{th}}$, and $13^{\text{th}}$
basepairs.}
\label{fig:NickHydro}
\end{figure}

\begin{figure}[!htb]
\centering
\includegraphics{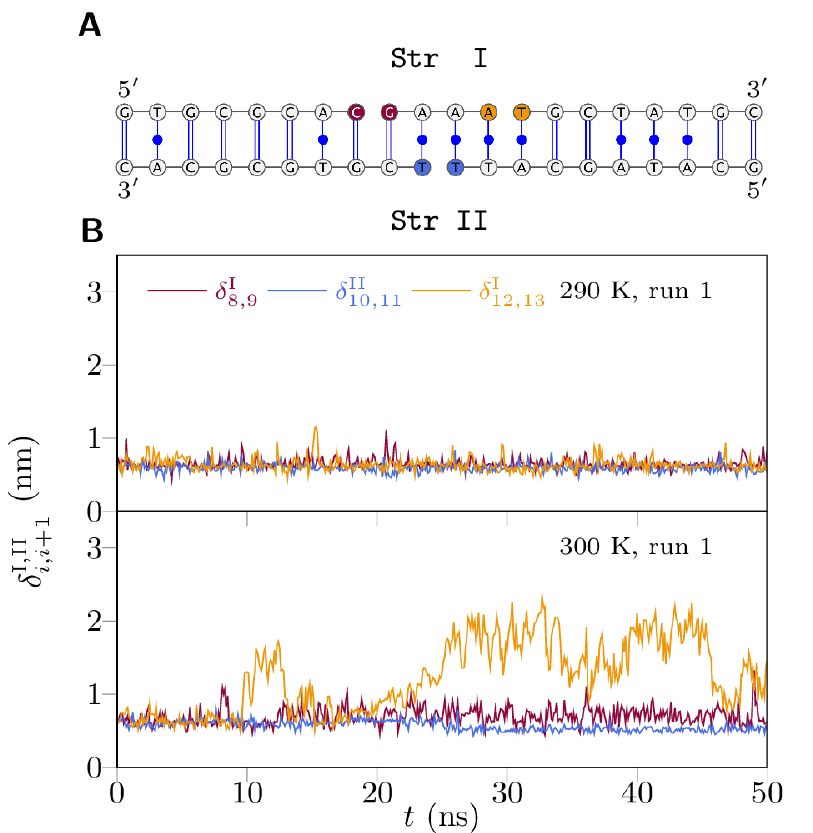}
\caption{\textbf{Temperature effects on nick unstacking.} At each temperature
of $290$, $300$, and $310$ K, four independent $50$ ns trajectories under the
spring constraint of $\left\{248.9;4.20\right\}$ were generated for
triple-nicked DNA. (\textbf{A}) An illustration of DNA with three nicks located
between the $8^\text{th}$ and $9^\text{th}$ basepairs on Strand \Rmnum{1}, the
$10^\text{th}$ and $11^\text{th}$ basepairs on Strand \Rmnum{2}, the
$12^\text{th}$ and $13^\text{th}$ basepairs on Strand \Rmnum{1}. The basepair
index, $i$, is counted from $5'$ end of Strand \Rmnum{1} as $1$ to $20$.
(\textbf{B}) The dynamics of interbase distances,
$\delta_{8,9}^\text{\Rmnum{1}}$, $\delta_{10,11}^\text{\Rmnum{2}}$, and
$\delta_{12,13}^\text{\Rmnum{1}}$, that straddling the nicks show clear
differences between $290$ and $300$ K, where lower temperature inhibits
unstacking. Note that only two out of twelve simulation dynamics are plotted
here as examples.}
\label{fig:TempC4Dyn}
\end{figure}

\clearpage
\section{Supporting Table}

\begin{table}[!htb]
\centering
\def\arraystretch{0.8}
\setlength\tabcolsep{0.2 cm}
\begin{tabular}{l c r r c r r c r r }
\toprule
 & & \multicolumn{2}{c}{Case \rmnum{1}} & & \multicolumn{2}{c}{Case \rmnum{2}} & & \multicolumn{2}{c}{Case \rmnum{3}} \\
\cline{3-4}
\cline{6-7}
\cline{9-10}
Bp step & & Tilt & Roll & & Tilt & Roll & & Tilt & Roll \\
\cline{1-10}
 01 \texttt{GT/AC} & &  0.51 & -3.81 & & -3.62 &  1.23 & &  2.73 &  2.68\\
 02 \texttt{TG/CA} & & -1.73 & -3.42 & & -2.27 &  3.09 & &  3.77 &  0.66\\
 03 \texttt{GC/GC} & & -3.35 & -1.83 & & -0.15 &  3.84 & &  3.49 & -1.59\\
 04 \texttt{CG/CG} & & -3.82 &  0.38 & &  2.05 &  3.23 & &  1.96 & -3.28\\
 05 \texttt{GC/GC} & & -2.92 &  2.47 & &  3.52 &  1.50 & & -0.23 & -3.82\\
 06 \texttt{CA/TG} & & -1.01 &  3.70 & &  3.75 & -0.75 & & -2.34 & -3.03\\
 07 \texttt{AC/GT} & &  1.26 &  3.62 & &  2.68 & -2.75 & & -3.65 & -1.16\\
 08 \texttt{CG/CG} & &  3.08 &  2.27 & &  0.65 & -3.78 & & -3.66 &  1.12\\
 09 \texttt{GA/TC} & &  3.83 &  0.12 & & -1.59 & -3.48 & & -2.40 &  2.98\\
 10 \texttt{AA/TT} & &  3.23 & -2.06 & & -3.29 & -1.96 & & -0.27 &  3.83\\
 11 \texttt{AA/TT} & &  1.50 & -3.53 & & -3.83 &  0.23 & &  1.92 &  3.31\\
 12 \texttt{AT/AT} & & -0.76 & -3.76 & & -3.00 &  2.37 & &  3.46 &  1.64\\
 13 \texttt{TG/CA} & & -2.76 & -2.66 & & -1.15 &  3.65 & &  3.79 & -0.61\\
 14 \texttt{GC/GC} & & -3.77 & -0.64 & &  1.13 &  3.66 & &  2.77 & -2.64\\
 15 \texttt{CT/AG} & & -3.48 &  1.62 & &  3.00 &  2.38 & &  0.79 & -3.75\\
 16 \texttt{TA/TA} & & -1.95 &  3.29 & &  3.82 &  0.27 & & -1.47 & -3.53\\
 17 \texttt{AT/AT} & &  0.26 &  3.82 & &  3.30 & -1.94 & & -3.21 & -2.08\\
 18 \texttt{TG/CA} & &  2.38 &  3.02 & &  1.63 & -3.46 & & -3.83 &  0.10\\
 19 \texttt{GC/GC} & &  3.64 &  1.13 & & -0.63 & -3.77 & & -3.10 &  2.25\\
\botrule
\end{tabular}
\caption{\textbf{Tilt and roll parameters for constructing directionally bent
initial conformations.} The rotational parameters, tilt and roll, describe the
relative rotational angles between consecutive basepair reference frames,
against $x$-axis and $y$-axis respectively. The constant
$\sqrt{\text{tilt}^2+\text{roll}^2}$ maintains a uniform bending, while their
systematic alternations alone helix produce a constant bending direction. These
three sets of tilt and roll angles (i.e., in units of degrees) generate DNA
initial conformations bending towards distinctive directions as shown in main
text Fig.~9, whose end-to-end vectors projected onto common $1^\text{st}$
basepair plane are evenly separated by $120^\circ$.}
\label{tab:TiltRoll}
\end{table}

\clearpage
\section{Supporting Methods: S\lowercase{imulation and analysis details}}
\subsection*{Unit cell preparation}

Before starting any simulation, a basic simulating unit (i.e., unit cell) was
properly constructed. Firstly, an initial atomic DNA structure with targeted
sequence and shape was generated using \texttt{X3DNA} \cite{Lu:2008kb}.
Secondly, this initial structure was centered within a minimal unit cell. Our
unit cell usually takes rhombic dodecahedron shape (i.e., $\sim 71\%$ of cubic
unit cell volume), whose inscribed sphere diameter equals the largest DNA
extension plus an additional $3.2$ nm for buffering. Next, this unit cell was
further prepared by filling the empty space with TIP3P water
\cite{Jorgensen:1983fl}, neutralizing the negative charges on DNA using sodium
counter-ions, and replacing some water molecules with sodium chloride to achieve
$150$ mM ionic strength. Lastly, it was finalized by energy minimization using
the steepest descent method to remove any energy unfavorable close contacts.

Based on this prepared unit cell, molecular trajectories were self-evolved
according to Newton's law of motion, given a set of initial velocities
(randomly sampled from Maxwell-Boltzmann distribution) and external
constraints, such as contractile springs for inducing bending. Before
collecting conformational evolutions, the unit cell was brought to correct
ensemble using $200$ ps velocity rescaling and $200$ ps Parrinello-Rahman
pressure coupling simulations
\cite{:/content/aip/journal/jcp/126/1/10.1063/1.2408420, Parrinello:1981it}.

\subsection*{Basepair coordinates}
Assigning a basepair coordinate to the group of thermally fluctuated atoms is
the key to bridge from MD raw trajectories to DNA macroscopic behaviors, such
as bending dynamics.

An ideal Watson-Crick basepair \cite{Clowney:1996kr} was fitted to each
observed instantaneous atomic arrangements during MD simulations by minimizing
the sum of squares of their residual errors. This least-square fitting was
implemented by Horn in 1987 \cite{Horn:1987hf} through finding a closed-form
solution of the ideal basepair absolute orientation against such instantaneous
atomic arrangements. A sketch of this ideal basepair coordinates is shown
(Fig.~2). For this \texttt{G$\equiv$C} Watson-Crick basepair, a right-handed
coordinate frame as described by Olson et al. \cite{Olson:2001cf} was attached
to it, with $\hat{x}_i$ pointing to the major groove, $\hat{y}_i$ pointing to
the backbone of the top strand, and $\hat{z}_i=\hat{x}_i \times \hat{y}_i$
describing the normal direction of the Watson-Crick basepair, where $i$ denotes
the $i^{\text{th}}$ basepair. This process was achieved using \texttt{X3DNA}
software during our analysis \cite{Lu:2003ug, Lu:2008kb}.

After this, some macroscopic configuration information was extrapolated using
local coordinates. For example, the bending angle between $i^\text{th}$ and
$\left( i+\Delta \right)^\text{th}$ basepairs, defined by $\theta_{i,i+\Delta}
= \cos^{-1} \left( \hat{z}_i \cdot \hat{z}_{i+\Delta} \right)$, where $i = 2,
3, \cdots, 19-\Delta$, were calculated for any instantaneous conformation of
DNA in some simulations.

\subsection*{Umbrella sampling}

During umbrella sampling, a series of springs indexed by $m$, each with a
finite equilibrium length of $l_m$ and fixed spring constant $ \kappa = 248.9$
pN/nm, were used to induce DNA bending (i.e., $\left\{248.9; l_m\right\}$). For
each $\left\{248.9; l_m\right\}$-constrained simulation, the biased
distribution of the distance fluctuation, $\rho_{\left\{ \kappa; l_m\right\}}
\left( d \right)$, was obtained. Theoretically, the regional unbiased ${\cal A}
\!\left(d \right)$ can be obtained by ${\cal A} \!\left(d \right)=-\beta^{-1}
\ln \rho_{\left\{ \kappa; l_m\right\}} \! \left(d \right)-\left( \kappa / 2
\right) \left(d -l_m \right)^2+{\cal A}_{\left\{ \kappa; l_m\right\}}$, where
${\cal A}_{\left\{ \kappa; l_m\right\}}$ is an undetermined shift. Here, the
unbiased distribution of the distance fluctuation $\rho \!\left(d \right)$ and
global $ {\cal A} \!\left(d \right)$ (i.e., reference to its global energy
minimal state) were obtained by weighted histogram analysis method using
\texttt{g\_wham} \cite{Kumar:1992bv, Hub:2010ex}, which optimizes the shifts to
minimize the statistical errors of $\sigma^2 \left(\rho \! \left(d
\right)\right)$ \cite{Torrie:1977hs}. During our analysis, $ {\cal A} \!\left(d
\right)$ was evaluated at $200$ discrete points for the each type of DNA, then
further smoothed by cubic spline interpolation, from which the continuous
force-distance curve could be obtained by $f \! \left( d\right) = - \partial
{\cal A}\!\left(d \right) / \partial d$.

\clearpage

\section{Supporting Discussion: A\lowercase{ctive bending experiment by}
S\lowercase{hroff et al.}}
In the experiment by Shroff et al., a nick-free $25$ bp dsDNA fragment was bent
by a $12$ nt ssDNA connected at the two dsDNA ends \cite{Shroff:2005cw}.
Assuming dsDNA is intact, its internal tension is expected to be around the
bucking transition force of $\sim 30$ pN. This corresponds to ssDNA separation
of $\sim 6$ nm based on phenomenological ssDNA force extension model
\cite{Cocco:2004fo}, similar to that between two points separated by $24$
basepair steps in a $64$ bp DNA minicircle with a planar circle conformation
(i.e., $L \sin \left( 24 \pi / 64 \right) / \pi \approx 6$ nm, where $L$ is the
contour length of the $64$ bp minicircle). However, the measured tension in the
ssDNA was shown to be $6 \pm 5$ pN, a few times smaller than the aforementioned
critical buckling force. Thus, the distance between the two dsDNA ends was
estimated to be only $< 4$ nm, which revealed DNA anomalous elastic responses.

\normalem